\documentclass[aps,prm,reprint, superscriptaddress]{revtex4-1}
\usepackage{graphicx}% Include figure files
\usepackage{dcolumn}% Align table columns on decimal point
\usepackage{bm}% bold math
\usepackage{amsmath}

\usepackage{color}
\usepackage{bm,graphicx,hyperref}
\hypersetup{%
  breaklinks = {true},
  citecolor = {blue},
  colorlinks = {true},
  linkcolor = {red},
}

\def\*#1{\mathbf{#1}}

\newcommand{\lth}{\ensuremath{\ell_\text{th}}}
\newcommand{\qth}{\ensuremath{q_\text{th}}}
\newcommand{\lp}{\ensuremath{\ell_\text{p}}}
\newcommand{\dd}{\text{d}}

\newcommand{\Ybare}{\ensuremath{Y}}
\newcommand{\YT}{\ensuremath{Y_\text{R}}}
\newcommand{\kappabare}{\ensuremath{\kappa}}
\newcommand{\kappaT}{\ensuremath{\kappa_\text{R}}}

\newcommand{\hcm}{\ensuremath{h_\text{CM}}}
\newcommand{\hM}{\ensuremath{h_\text{M}}}
\newcommand{\kBT}{\ensuremath{k_{\text B}T}}
\newcommand{\Ahcm}{\ensuremath{A_{h_{\rm CM}}}}
\newcommand{\tauD}{\ensuremath{\tau_{\rm damp}}}
\newcommand{\tauoT}{\ensuremath{\tau_{\rm o}^{\rm R}}}
\newcommand{\taubT}{\ensuremath{\tau_{\rm b}^{\rm R}}}
\newcommand{\woT}{\ensuremath{\omega_{\rm o}^{\rm R}}}
\newcommand{\wo}{\ensuremath{\omega_{\rm o}}}
\newcommand{\wD}{\ensuremath{\omega_{\rm D}}}

\newcommand{\Lex}{\ensuremath{\mathcal{L}}}
\newcommand{\Lbath}{\ensuremath{\mathcal{L}_\text{bath}}}
\newcommand{\Lbeam}{\ensuremath{\mathcal{L}_\text{beam}}}

\usepackage{setspace}
\begin{document}

\title{Vibrations and tunneling of strained nanoribbons at finite temperature}

\author{Paul Z. Hanakata}
\affiliation{Department of Physics, Harvard University, Cambridge, Massachusetts 02138, USA}
\email{paul.hanakata@gmail.com}

\author{Sourav S. Bhabesh}
\thanks{Work completed prior to joining AWS.}
\affiliation{Amazon Web Services (AWS), Washington DC Metro Area, USA}

\author{David Yllanes}
\affiliation{Chan Zuckerberg Biohub, San Francisco, CA 94158, USA}
\affiliation{Instituto de Biocomputación y Física de Sistemas Complejos (BIFI), 50018 Zaragoza, Spain}

\author{David R. Nelson}
\affiliation{Department of Physics, Harvard University, Cambridge, Massachusetts 02138, USA}

\author{Mark J. Bowick}
\affiliation{Kavli Institute for Theoretical Physics, University of California, Santa Barbara, CA 93106, USA}
\email{markbowick@ucsb.edu}

\date{\today}
%\linenumbers
\begin{abstract}
Crystalline sheets (e.g., graphene and transition metal dichalcogenides) liberated from a substrate are a paradigm for materials at criticality because flexural phonons can fluctuate into the third dimension. Although studies of static critical behaviors (e.g., the scale-dependent elastic constants) are plentiful, investigations of dynamics remain limited. Here, we use molecular dynamics to study the time dependence of the midpoint (the height center-of-mass) of doubly clamped nanoribbons, as prototypical graphene resonators, under a wide range of temperature and strain conditions. By treating the ribbon midpoint as a Brownian particle confined to a nonlinear potential (which assumes a double-well shape beyond the buckling transition), we formulate an effective theory describing the ribbon's tunneling rate across the two wells and its oscillations inside a given well. We find that, for nanoribbbons compressed above the Euler buckling point and thermalized above a temperature at which the non-linear effects due to thermal fluctuations become significant, the exponential term (the ratio between energy barrier and temperature) depends only on the geometry, but not the temperature, unlike the usual Arrhenius behavior. Moreover, we find that the natural oscillation time for small strain shows a non-trivial scaling $\tau_{\rm o}\sim L_0^{\,z}T^{-\eta/4}$, with $L_0$ being the ribbon length, $z=2-\eta/2$ being the dynamic critical exponent, $\eta=0.8$ being the scaling exponent describing scale-dependent elastic constants, and $T$ being the temperature. These unusual scale- and temperature-dependent dynamics thus exhibit dynamic criticality and could be exploited in the development of graphene-based nanoactuators. 
\end{abstract}

\maketitle
\section{Introduction}
In the last decade there has been growing interest in utilizing mechanical instabilities in thin materials to design smart materials with desired functionalities, from grasping~\cite{dias-sm-48-9087-2017, yang2021grasping} and shape morphing~\cite{tao2021morphing, liu2021frustrating} to locomotion~\cite{nagarkar2021elastic, lee2022buckling}. Using membranes, such as thin sheets, as a building block (say an oscillator) for soft-robotic applications is appealing because thin sheets are flexible and can be controlled with minimal and simple actuation. The buckling instability, which sets in for sufficiently large F\"oppl-von K\'arm\'an number ${\rm vK}=\frac{YA}{\kappa}$, where $Y$ is the 2D Young's modulus, $A$ is a characteristic ribbon area, and $\kappa$ the bending rigidity, is an important mechanism for such actuation. This simple principle has been successfully applied to a wide range of materials and system sizes, ranging from meter-sized satellites to nanoactuators~\cite{miura-1985, shyu-NatMat-14-785-2015, dias-sm-48-9087-2017, callens-MT-2017, yang2021grasping, melancon2021multistable, nagarkar2021elastic, lee2022buckling}. Very recently, there has been success in applying instability mechanisms to control actuator movements in low-noise environment, for example in a centimeter-sized buckling-sheet oscillator~\cite{nagarkar2021elastic, lee2022buckling}. It remains to be seen, however, if similar principles apply in a more noisy environment with, for example, strong thermal fluctuations. 

The mechanical response and energy dissipation of micro- and nanoscale oscillators have long been studied~\cite{Panov1985, Dykman2012}. Graphene and other 2D-materials-based nanoresonators, commonly in a double-clamped geometry, have been studied extensively. They exhibit remarkable properties compared to their bulk counterparts, including tunability over a wide frequency range,  kilo- to terahertz, and a very high quality factor~\cite{bunch2007electromechanical, zande2010large, chen2009performance, aleman2013polymer, Chen2017, jiang2015review, akinwande2017review, blaikie2019fast, miller2017shape}. Exciting though these features are, precise control of the thermal dynamics of these atomically thin materials remains a challenge and is crucial for building, say, soft robots~\cite{nagarkar2021elastic, lee2022buckling}. Nevertheless, nature has shown us that micro- to nanosized biological ``robots," such as kinesins and other molecular motors, do exist at biologically relevant temperatures.

One of the main challenges in building 2D-materials-based robots or actuators is that height corrugations due to thermal fluctuations~\cite{roldan-PRB-83-174104-2011, kovsmrlj-PRB-93-125431-2016, bowick-PRB-95-104109-2017}, impurities~\cite{plummer2020buckling, hanakata-PRL-128-075902-2022, phn-cylinder-2022}, or quenched disorder~\cite{kovsmrlj2013mechanical}, alter the mechanics significantly at large distances---similar to how a wrinkled paper sheet can bear its own weight while a pristine sheet sags. Indeed the bending rigidity of a micron-sized graphene ribbon has been observed experimentally to exhibit a striking $\sim$ 4000-fold increase at room temperature relative to its zero-temperature value, demonstrating the non-trivial mechanics of nanomaterials~\cite{blees2015graphene}. Because the mechanical properties are scale-dependent, which may complicate dynamics, scaling up a micron-sized robot based on graphene nanoribbons or nanotubes requires new design principles. Moreover, while fundamental studies of electronic, optical, and mechanical properties of graphene and other 2D materials are numerous~\cite{neto2009electronic, wang2012electronics, novoselov20162d, akinwande2017review, RevModPhys.93.011001}, there has been much less focus on their dynamical behavior~\cite{ackerman2016anomalous,granato2022dynamic}, in particular dynamical critical exponents that relate time scales to length scales.  

As mechanical properties play an important role in determining the dynamics, such as underdamped or overdamped oscillations, we develop here a framework, motivated by extensive molecular dynamics simulations, to analyze the dynamics of nanoribbons over a wide range of temperatures and strains. We focus specifically on doubly clamped ribbons as one of the most common geometries for nanoelectromechanical systems (NEMS). In contrast to recent work~\cite{Dou2015}, in which thermal effects are neglected while designing clamped resonators, we propose a simple geometric tunability that exploits thermal fluctuations as a means of studying anharmonic effects and dynamics.

We will demonstrate that the dynamics of nanoribbons has two distinct behaviors at and above the temperature at which the thermal renormalization of elastic constants sets in. In Sec.~\ref{sec:model} we introduce a simple computational model of nanoribbons mimicking 2D materials such as graphene. We first show how the height fluctuations change with strain in Sec.~\ref{sec:renormalized-mechanics}, demonstrating the scale-dependent mechanics with simulations. In Sec.~\ref{sec:coarse-graining}, we propose an effective free energy of the strained nanoribbon and present molecular dynamics results of the motion of nanoribbons under various strain conditions. We then develop a phenomenological model treating the midpoint of a ribbon as a Brownian particle with damping confined in a nonlinear potential with both single and double wells to understand the dynamics of nanoribbons under compression (Sec.~\ref{sec:compressed-ribbon}) and stretching (Sec.~\ref{sec:stretched-ribbon}). In each respective section we present molecular dynamics simulations checking our theoretical predictions. 

We find that the escape time of the midpoint, which characterizes the inverse of the ribbon flipping rate, of a compressed ribbon at sufficiently high temperatures is approximately temperature-independent and solely governed by the geometry, unlike the usual Arrhenius behavior. At sufficiently high temperatures, where renormalization becomes important, the characteristic escape time scales with system size as $\tau_p\sim L_0^{\,4-\eta}$, in the high-damping regime, and independent of system size in the low-damping regime, with $\eta \approx 0.8$  the exponent controlling the scale-dependent bending rigidity and  $L_0$ the ribbon length. For a slightly stretched or relaxed ribbon we find that the natural oscillation time (oscillation time inside a minima) scales as $\tau_{\rm o}\sim L_0^{(2-\eta/2)}T^{-\eta/4}$, which has no analog in standard mechanical resonators. In the language of dynamic critical phenomena~\cite{hohenberg1977}, we have a dynamic critical exponent $z=2-\eta/2$ for relaxed ribbons, and $z=1$ for ribbon under tension, consistent with Van Hove, with no singularities in the transport coefficients.  

We conclude by discussing future prospects, including further investigation of the connection between the dynamical critical exponent $z$ and the static exponent $\eta$ using  finite-size scaling,  as well as incorporating an attractive substrate in the numerical simulations to capture energy losses present in certain experiments. 
\section{The Model}
\label{sec:model}
Similar to a number of previous studies~\cite{seung-PRA-38-1005-1988, Bowick1996, bowick-PRB-95-104109-2017, yllanes-NatCom-8-1-2017, hanakata-EML-44-101270-2020}, we simulate ribbons discretized on a equilateral triangular lattice. The ribbon is comprised of $N_x\times N_y=100\times25$ nodes with rest (zero-temperature) length $L_0\sim100a$ and width $W_0\sim20a$. To model a doubly clamped ribbon, the nodes in the two rows at each end are held fixed. We use a standard coarse-grained model~\cite{seung-PRA-38-1005-1988} to compute the total energy of the ribbon. Each node is connected by a harmonic spring with a rest length of $a$. The bending energy is computed using the  dihedral interaction between the normals. The total energy is given by
\begin{equation}
  E=\frac{k}{2}\sum_{\langle i,j\rangle}||\mathbf{r}_{i}-\mathbf{r}_j|-a|^2+\hat{\kappa}\sum_{\langle\alpha, \beta\rangle}(1-\mathbf{n}_{\alpha}\cdot\mathbf{n}_{\beta})
\end{equation}
where $k$ is the harmonic spring constant and $\hat{\kappa}$ is the microscopic bending rigidity. The first sum is over neighboring nodes and the second sum is over neighboring triangles. The continuum limit yields $\kappabare=\sqrt{3}\hat{\kappa}/2$ for the bare (zero-temperature) continuum bending rigidity and $\Ybare=\sqrt{2}k/3$ for the bare continuum 2D Young's modulus~\cite{seung-PRA-38-1005-1988}. Following~\cite{bowick-PRB-95-104109-2017, yllanes-NatCom-8-1-2017}, we set $k=1440\hat{\kappa}/a^2$ so that the F\"oppl-von K\'arm\'an number vK = $\Ybare W_0L_0/\kappabare\sim10^6$ is experimentally realistic. This coarse-grained model has been widely used to model atomically thin materials such as graphene and MoS$_2$ and successfully captures mechanical and thermal response~\cite{Bowick1996, zhang-JMPS-67-2-2014, morshedifard-JMPS-149-104296-2021, hanakata-EML-44-101270-2020, bowick-PRB-95-104109-2017, wan2017thermal} consistent with those found in simulations with more sophisticated atomistic potentials~\cite{fasolino2007intrinsic, roldan-PRB-83-174104-2011, jiang2014buckling, hanakata-Nanoscale-8-458-2016, ahmadpoor2017thermal, hanakata-PRR-2-042006-2020, hanakata-PRL-128-075902-2022}. 

The molecular dynamics (MD) simulations are performed with the HOOMD-blue package~\cite{anderson2020hoomd} within the $NVT$ ensemble (fixed number of particles $N$, volume $V$ and temperature $T$) with an integration time step of $dt=0.001\tau_{\rm MD}$, where $\tau_{\rm MD}=\sqrt{\mathcal{M}\mathcal{D}^2/\mathcal{E}}$ is the MD unit of time and $\mathcal{M, D, E}$ are the fundamental units of mass, distance, and energy. For graphene parameters, $\tau_{\rm MD}\sim1$~ps. Temperature is controlled every $\tau_T=0.2\tau_{\rm MD}$ via the Nos\'{e}-Hoover thermostat~\cite{martyna1994constant}.  For systems clamped at compressive strains below critical buckling, we run a total of $10^7$ steps and discard 50\% of the data for thermal equilibration. Above the critical buckling the relaxation time increases significantly, and therefore we run a total of $10^8$ steps and discard the first 20\% of the data for thermal equilibration. Snapshots are taken every 10,000 steps or equivalently $10\tau_{\rm MD}$. HOOMD scripts and analysis codes used in this study are available at \url{https://github.com/phanakata/statistical-mechanics-of-thin-materials/}. All simulation data will be reported in natural MD units $\mathcal{D}=\mathcal{M}=1$, $\kBT$ in units of $\hat{\kappa}$ and time in $\tau_{\rm MD}$. Temperature is reported as the ratio of the ribbon width $W_0$ to the thermal length~\cite{kovsmrlj-PRB-93-125431-2016} $\lth=\sqrt{\frac{64\pi^3\kappa^2_0}{3k_{\rm B}T\,Y_0}}$, as explained below.

To study ribbon dynamics over a wide temperature range we vary the ratio of temperature to microscopic bending rigidity  ${\kBT}/\hat{\kappa}$ over a wide range, from $10^{-1}$ to $10^{-5}$, while keeping $k=1440\hat{\kappa}/a^2$ and the preferred bond length constant at $a=1$. Strains $\epsilon$ are applied by clamping the two ends of the ribbon at different lengths $L_\epsilon$. Thermal fluctuations lead to a reduced projected length of the unstrained ribbon, $L_{\rm relax}$, relative to the  unstrained length at zero temperature $L_0$.  The relaxed length $L_{\rm relax}$ is determined by the vanishing of the average longitudinal stress $\langle \sigma_{xx}\rangle$~\cite{hanakata-EML-44-101270-2020}. The compressive strain $\epsilon=(1-L_\epsilon/L_{\rm relax})$ is measured relative to the unstrained ribbon with clamping at $L_{\rm relax}$. 
\section{Height profile of deformed ribbons}
\label{sec:renormalized-mechanics}
\begin{figure}[h]
\centering
\includegraphics[width=8.6cm]{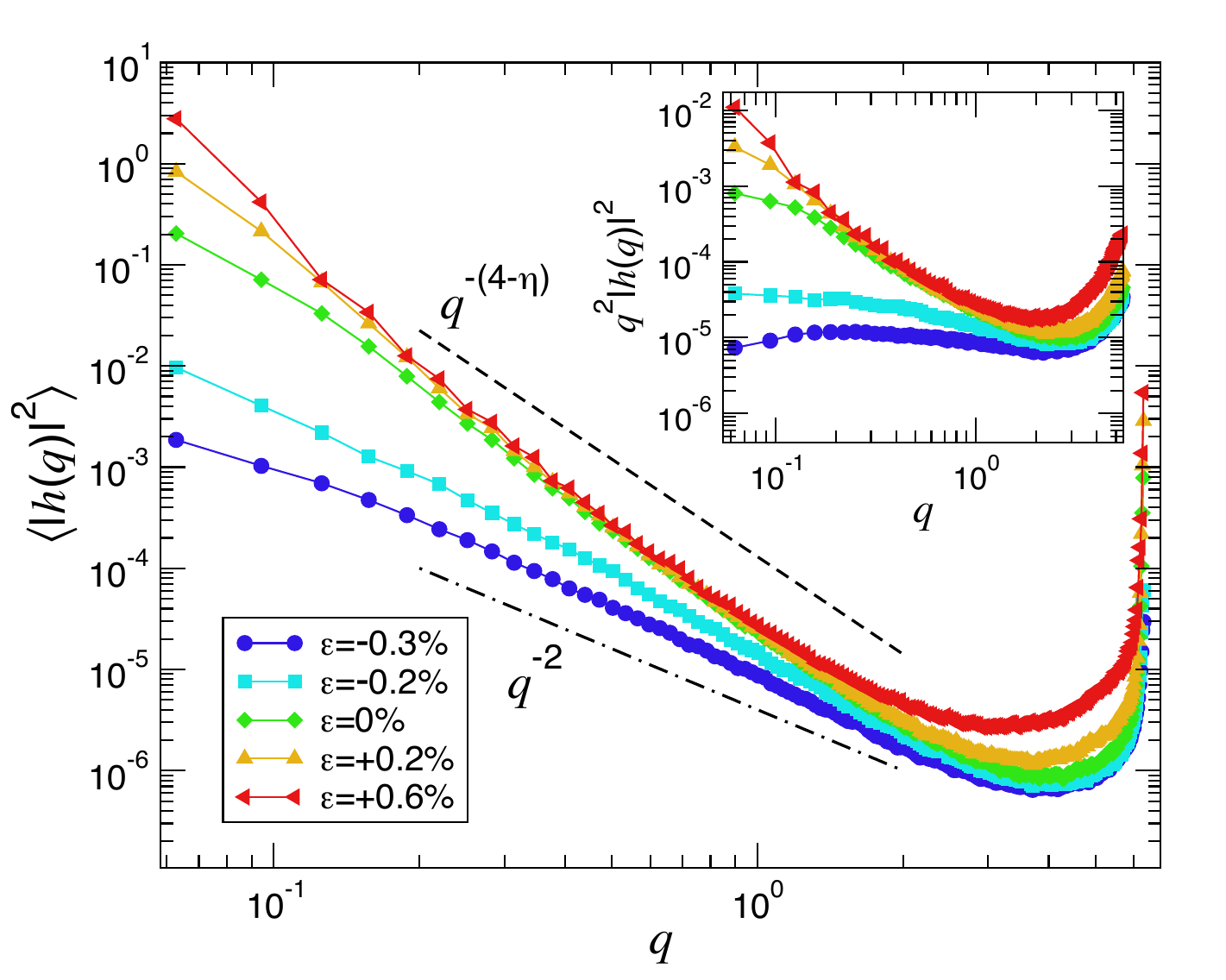}
\caption{The height fluctuations $\langle |h(q)|^2\rangle$ as a function of wavevector $q$ for a ribbon clamped at different strains, both extensional ($\epsilon<0$) and compressional ($\epsilon>0$), $\epsilon=[-0.3\%, -0.2\%, 0\%, +0.2\%, +0.6\%]$. $h(q)$ is obtained from the  Fourier transform $h(q)=\frac{1}{A_0}\int e^{i(q_x\,x + q_y\, y)}h(x,y)\,dx\,dy$, where $q_x, q_y$ are wavevectors and $A_0=W_0\times L_0$ is the area of the unstrained (rest) ribbon at zero temperature. We use a finite number of $q$ modes ranging from $|q_{\rm min}|=\pi/L$ to $|q_{\rm max}|=2\pi/a$, with an increment of $\Delta q=\pi/L$, and set $q_y=0$. Temperature is set at  $k_\text{B}T=0.05\hat{\kappa}$, which corresponds to $W_0/\lth=8.5$, so thermal renormalizations are strong. For stretched ribbons ($\epsilon<0$), $\langle |h(q)|^2\rangle$ is proportional to $q^{-2}$ for a wide range of $q$. For unstrained ($\epsilon=0$) and compressed ribbons ($\epsilon>\epsilon_c$, above the thermalized Euler buckling threshold), $\langle |h(q)|^2\rangle$ scales like $q^{-(4-\eta)}$ with $\eta \approx 0.8$. The black dashed line and the black dotted-dashed line show $q^{-(4-\eta)}$ and $q^{-2}$ scaling, respectively. The inset shows $q^2\langle |h(q)|^2\rangle$ versus $q$ to more clearly bring out the $q^{-2}$ dependence of stretched ribbons.}
\label{fig:hq2}
\end{figure}
Before discussing the dynamics of thermalized nanoribbons, we first probe the effects of strains on static properties. To lowest order in the height field $h(x,y)$, in-plane displacement $\mathbf{u}(x,y)$ and their gradients, the elastic energy of a membrane under a spatially uniform uniaxial compression along the $x$ direction, $\sigma_{xx}$, can be written in the continuum limit as~\cite{roldan-PRB-83-174104-2011, kovsmrlj-PRB-93-125431-2016}
\begin{align}
    G[\mathbf{u}, h]=&\frac{1}{2}\int dx\,dy \left[\kappabare\left( \nabla^2 h\right)^2+2\mu u_{ij}^2+\lambda u_{kk}^2\right]\nonumber\\ &- \int dx\,dy\, \sigma_{xx}(\partial_x u_x),
\end{align}
where $u_{ij}\approx(\partial_i u_j+\partial_j u_i)/2+\partial_i h \partial_j h$ is the nonlinear strain tensor, $\kappabare$ is the bare continuum bending rigidity and $\mu$ and $\lambda$ are the Lam\'{e} coefficients. By tracing out the in-plane degrees of freedom, the effective free energy can be written in terms of out-of-plane flexural phonon deformation field $h(x, y)$~\cite{kovsmrlj-PRB-93-125431-2016}
\begin{align}
    G_{\rm eff}[h]=&\int dx\,dy \left[\frac{\kappabare}{2}\left(\nabla^2 h\right)^2+\frac{\Ybare}{8}\left(P_{ij}^T(\partial_i h)(\partial_j h)\right)^2\right]\nonumber\\&-\int dx\,dy\sigma_{xx}(\partial_x h)^2,
\end{align}
where $\Ybare=4\mu(\mu+\lambda)/(2\mu+\lambda)$ is the bare 2D Young's modulus and $P_{ij}^T=\delta_{ij}-\partial_i\partial_j/\nabla^2$ is the transverse projection operator. Within the harmonic approximation, the spectrum of the height-height correlation function of a tensionless sheet is $\langle |h(q)|^2\rangle=\kBT/(A_0\kappabare q^4)$, where $A_0=L_0\times W_0$ is the undeformed sheet area, and $h(\mathbf{q})\equiv\frac{1}{A_0}\int dx\,dy\,e^{-(q_x\, x+q_y\,y)}h(x,y)$ is the Fourier transform of $h(x,y)$. At low-temperature ($\kBT/\kappa\ll1$) a perturbative calculation shows that the bending rigidity is renormalized by thermal fluctuations in the form $\kappa(\mathbf{q})=\kappa_0+\frac{Y_0\kBT}{\kappabare}I(\mathbf{q})$, where $\mathbf{q}$ is the  wavevector and $I(\mathbf{q})$ is a momentum integral that scales as $q^{-2}$ for $q\rightarrow 0$~\cite{Statmech}. The relative perturbative correction is of order one above a fundamental length scale  $\lth\sim\sqrt{\kappabare/(\Ybare\kBT)}$~\cite{Peliti1987, Statmech}. At and above $\lth$ thermal fluctuations lead to  scale-dependent mechanical moduli and non-trivial departures from the expected zero-temperature mechanical behavior. Within a renormalization group treatment, the spectrum of the height-height correlation function of a ribbon under uniaxial compression is given by~\cite{kovsmrlj-PRB-93-125431-2016} 
\begin{equation}
    \langle |h(q)|^2\rangle=\frac{\kBT}{A_0(\kappaT(q)q^4-\sigma_{xx}q_x^2)},
\label{eq:hq2}
\end{equation}
where $\sigma_{xx}\simeq \YT\epsilon$ is the positive compressive stress. The scale-dependent renormalized bending rigidity, $\kappaT(q)$, and 2D Young's modulus, $\YT(q)$, are given by~\cite{kovsmrlj-PRB-93-125431-2016, Statmech} 
\begin{align}
\label{eq:relastics_1}
\kappaT(q) &\sim
  \begin{cases}
    \kappabare     & \quad \text{if } q\gg \qth\\
    \kappabare\left(q/\qth\right)^{-\eta}  & \quad \text{if } q\ll \qth
  \end{cases}\\
  \YT(q) &\sim
  \begin{cases}
    \Ybare     & \quad \text{if } q\gg \qth\\
    \Ybare\left(q/\qth\right)^{\eta_u}  & \quad \text{if } q\ll \qth
  \end{cases}
  \label{eq:relastics_2}
\end{align}
where $\eta$ and $\eta_u$ are scaling exponents and $\qth\equiv2\pi/\lth=\sqrt{\frac{3k_{\rm B}T\,Y_0}{16\pi\kappa^2_0}}$ is the wavevector below which renormalization becomes important~\cite{Peliti1987}. Theoretical estimates~\cite{Peliti1987, Aronovitz1988, doussal-PRL-69-1209-1992} of the scaling exponents give $\eta\approx0.8-0.85$ and $\eta_u\approx0.2-0.4$, and have been confirmed by height-height correlation measurements in Monte Carlo~\cite{Bowick1996, fasolino2007intrinsic, roldan-PRB-83-174104-2011, troster-PRB-87-104112-2013} and in molecular dynamics simulations~\cite{zhang1993scaling, bowick-PRB-95-104109-2017, hanakata-PRL-128-075902-2022}, as well as more recently by stress-strain curve measurements~\cite{morshedifard-JMPS-149-104296-2021, hanakata-EML-44-101270-2020}. 

Eq.~\ref{eq:hq2} indicates that height fluctuations are suppressed when stretching ($\sigma_{xx}<0$) is applied. For sufficiently large stretching $|\epsilon|\gg \kappaT/(\YT q^2)$ the $q^{-2}$ behavior in $\langle |h(q)|^2\rangle$ should dominate. Equivalently, for small wavevectors, $q\ll \sqrt{|\epsilon|Y(q)/\kappaT(q)}$, $\langle |h(q)|^2\rangle$ should switch from a $q^{-(4-\eta)}$ or $q^{-4}$ dependence to a $q^{-2}$ fall-off. 
Fig.~\ref{fig:hq2} shows the spectrum of the height-height correlation $\langle |h(q)|^2\rangle$ obtained from MD simulations as a function of wavevector $q$ for five different strains, both compressional $\epsilon>0$ and extensional $\epsilon<0$, $\epsilon=[-0.3\%, -0.2\%, 0\%, +0.2\%, +0.6\%]$. Here we show a system at a sufficiently high temperature, $\kBT/\hat{\kappa}=0.05$ ($W_0/\lth=8.5$), where thermal fluctuations are significant. The thermalized critical Euler buckling strain for this particular system is $\epsilon_c=0.05\%$. In the unstrained case, we see that $\langle |h(q)|^2\rangle \sim q^{-(4-\eta)}$, with $\eta \approx 0.8$, as expected~\cite{kovsmrlj-PRB-93-125431-2016, morshedifard-JMPS-149-104296-2021}, for a wide range of $q$.  For stretched ribbons, in contrast, $\langle|h(q)|^2\rangle$ scales more like   $q^{-2}$. This is better seen in the plot of $q^2\langle |h(q)|^2\rangle$ in the inset of Fig.~\ref{fig:hq2}. While stretching ($\epsilon<0$) suppresses height fluctuations, sufficiently large compression drives buckling and consequently $\langle |h(q)|^2\rangle$ of a compressed ribbon is elevated relative to the unstrained case. These strain-induced modifications of  static properties have also been observed in the normal-normal correlation function of graphene under isotropic deformation~\cite{roldan-PRB-83-174104-2011}. 
\section{Mean-field approximation to ribbon midpoint energetics}
\label{sec:coarse-graining}
\begin{figure}[h]
\centering
\includegraphics[width=8.6cm]{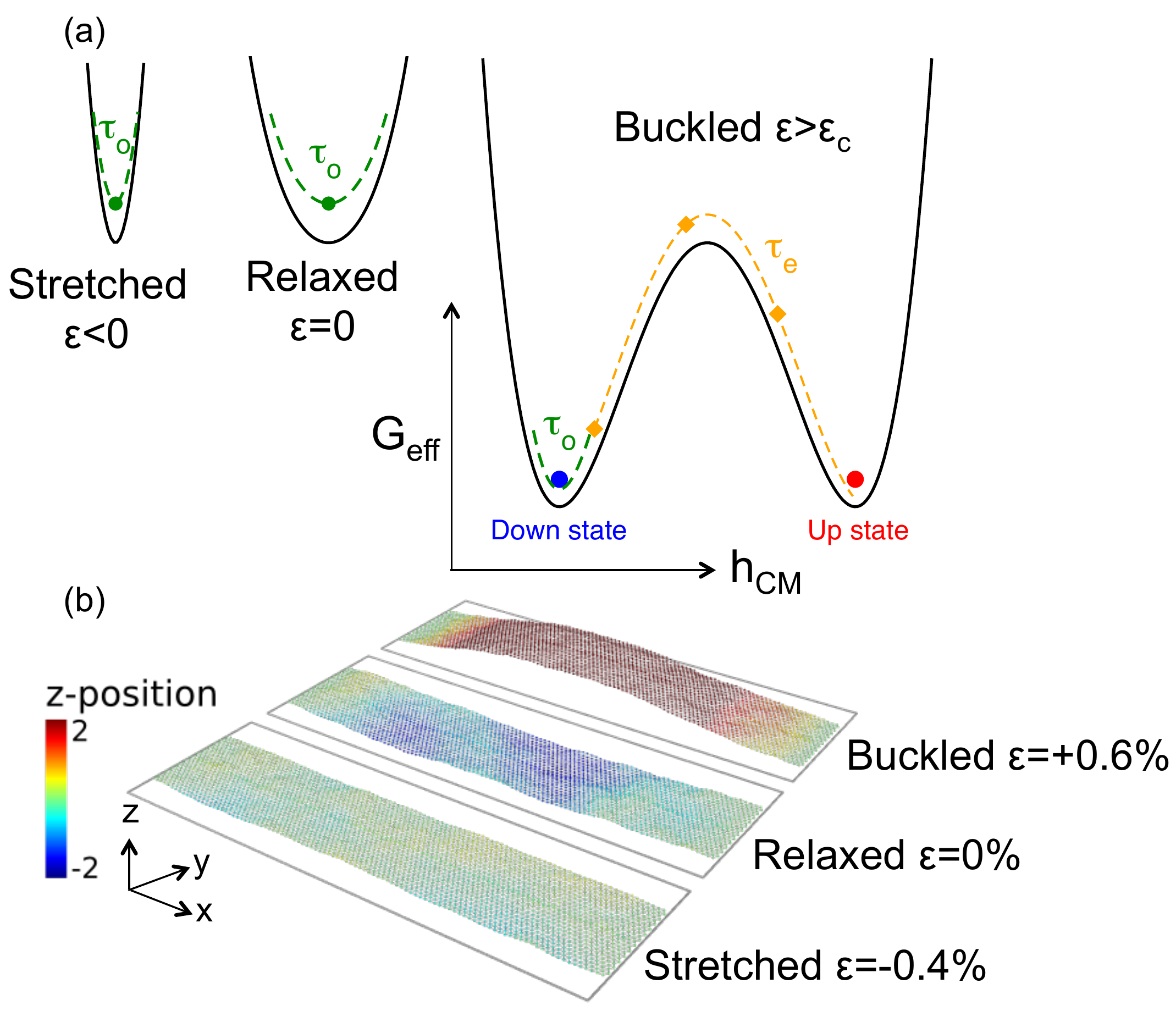}
\caption{(a) Schematics of the mean-field Gibbs free energy $G_{\rm eff}$ as a function of height center-of-mass, $\hcm$,  for a ribbon under stretched, unstrained, and buckled conditions. In each well the center of mass oscillates with a period of $\tau_{\rm o}=2\pi\sqrt{\frac{M}{k^{\rm eff}}}$, where $M$ is the ribbon mass, $k^{\rm eff}=\left.\frac{d^2 G_{\rm eff}}{d\hcm^2}\right|_{\hcm=h^*_\text{CM}}$ is the effective spring constant, and  $h^*_\text{CM}$ is the $\hcm$ where $G_{\rm eff}$ is at a minimum. (b) Representative configurations of a ribbon corresponding to three different compressive strains: $\epsilon=-0.4\%$ (stretched), $\epsilon=0\%$ (unstrained), and $\epsilon=+0.6\%$ (buckled). Recall that the critical strain for compressive buckling under these conditions is quite small, $\epsilon_c=0.05\%$. The color represents the $z$-position of a node scaled to the range $-2a$ to $+2a$. Positions are visualized using OVITO software~\cite{ovito}.}
\label{fig:schematics_hprofile}
\end{figure}
We turn now to a simplified model of the dynamics of the ribbon center of mass, which is related to the fundamental mode of a doubly clamped ribbon. We simplify by coarse-graining over the short-scale fluctuations along the $x$ and $y$ directions. Specifically, we assume that the height profile is constant along the $y$ direction in  Fig.~\ref{fig:schematics_hprofile}(b). For a ribbon of width $W_0$, this approach  effectively treats the ribbon as a one-dimensional object but with modified $W_0$-dependent elastic constants. 
By integrating out the in-plane phonons, the effective Gibbs free energy becomes~\cite{hanakata-EML-44-101270-2020}
\begin{equation}
\begin{split}
 G_{\rm eff}[h]&=\frac{\kappa W_0}{2}\int_{-L_\epsilon/2}^{L_\epsilon/2} dx\left(\frac{d^2h}{d x^2}\right)^2\\
&\quad +\frac{YW_0}{2L_{\epsilon}}\left[\int_{-L_\epsilon/2}^{L_\epsilon/2}dx\frac{1}{2}\left(\frac{dh}{dx}\right)^2\right]^2\\
&\quad -\frac{F}{2}\int_{-L_\epsilon/2}^{L_\epsilon/2}d x\left(\frac{d h}{d x}\right)^2 dx+ G^{\rm pre}[\overline{\Delta L}],
\end{split}
\label{eq:G}
\end{equation}
where $L_\epsilon$ is the projected ribbon length corresponding to the strain $\epsilon$ and $G^{\rm pre}$ is the total prestress elastic energy stored during compression before buckling. $G^{\rm pre}$ is independent of the ribbon height profile and can be dropped. Within the mean-field approximation, the ribbon height is assumed to be smooth over scales larger than the thermal length $\lth$ and double-clamped boundary condition is implemented. These two conditions can be approximated by a profile $h(x)=\frac{\hM}{2}\left[1+\cos\left(\frac{2\pi x}{L_{\epsilon}}\right)\right]$. Upon using this height as an ansatz we obtain the effective Gibbs free energy from Eq.~\ref{eq:G}~\cite{hanakata-EML-44-101270-2020} 
\begin{equation}
    G_{\rm eff}[\hM]=\frac{\pi^2\Ybare W_0}{4L_\epsilon}(\epsilon_c-\epsilon)\hM^2+\frac{\pi^4\Ybare W_0}{32L_\epsilon^3}\hM^4, 
\label{eq:GhA}
\end{equation}
where $\epsilon_c=\frac{4\pi^2\kappabare}{\Ybare L_{\epsilon_c}^2}$ is the critical strain for Euler buckling and $L_{\epsilon_c}$ the associated projected length. Although this energy resembles the Landau theory of a critical point, note that $\epsilon_c$ (the analog of a critical temperature) depends on the system size. For $\epsilon>\epsilon_c$, there are two stable minima at $\hM=\pm\frac{2L_{\epsilon_\text{c}}}{\pi}\sqrt{\epsilon-\frac{4\kappabare\pi^2}{\Ybare L_{\epsilon_\text{c}}^2}}$ and one unstable point at $\hM=0$, whereas for $\epsilon\leq\epsilon_c$ there is one stable minimum at $\hM=0$ (see Fig.~\ref{fig:schematics_hprofile}(a)). To relate this result to simulations, we use the center-of-mass midpoint $\hcm=\frac{1}{N}\sum_i z_i$ as a measure  of the aggregate collective motion of all nodes. This simplification effectively treats the ribbon as a Brownian particle confined to a nonlinear potential. Henceforth we will write Eq.~\ref{eq:GhA} and other derived quantities in terms of $\hcm$ using   $\hcm^2\equiv\left(\frac{1}{L_\epsilon}\int^{L_\epsilon/2}_{-L_\epsilon/2}h\
  \dd x\right)^2=\frac14{\hM^2}$~\cite{hanakata-EML-44-101270-2020}.  

\begin{figure}[h]
\centering
\includegraphics[width=8.6cm]{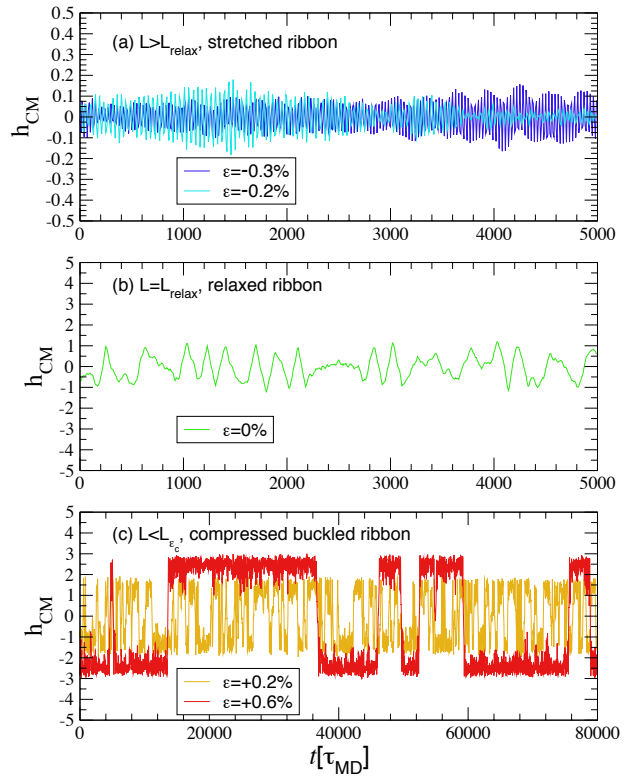}
\caption{Midpoint trajectory $\hcm(t)$ of a ribbon under (a) stretched $\epsilon=[-0.3\%, -0.2\%]$, (b) unstrained $\epsilon=0\%$, and (c) buckled $\epsilon=[+0.2\%, +0.6\%]$ conditions at a fixed $W_0/\lth=8.5$ ($\kBT=0.05\hat{\kappa}$). The time $t$ is in units of the MD time unit $\tau_{\rm MD}$. For clarity the time domain is chosen differently in (c). For stretched ribbons (a) $\hcm$ oscillates rapidly about the zero plane with small amplitude. For unstrained ribbons (b) the oscillation period increases and is irregular. Well beyond the thermalized Euler buckling point ribbons stay buckled either above or below the zero-plane for many oscillations before switching to the other local minimum (up to down state and vice versa). In (c) we see a dramatic increase in residence time with increasing compressive strain. In a local minimum, $\hcm$ fluctuates with a shorter period and with a smaller amplitude relative to the tunneling (fluctuation over a barrier) dynamics, indicating that $\hcm$ oscillates inside the local minimum for many periods before escaping over the potential barrier.}
\label{fig:hcm-t}
\end{figure}

Eq.~\ref{eq:GhA} reveals that when $\epsilon\ll\epsilon_c$ the non-linearity can be neglected, and for small height deflections, $\hcm$ is expected to oscillate in  a harmonic potential with a period $\tau_{\rm o}=2\pi/\wo$, where we expect $\wo$ is related to the total ribbon mass $M$ by $\wo=\sqrt{k^{\rm eff}/M}$ and $k^{\rm eff}=\left.\frac{d^2 G_{\rm eff}}{d\hcm^2}\right|_{\hcm=h^*_\text{CM}}$, where $h^*_\text{CM}$ is the minimum shown on the right side of Fig.~\ref{fig:schematics_hprofile}(a). From MD simulations, we indeed find that the $\hcm$ of a ribbon stretched at $\epsilon=[-0.2\%, -0.3\%]$ oscillates with a sinusoidal-like function around zero, as shown in Fig.~\ref{fig:hcm-t}(a). For the unstrained case, shown in Fig.~\ref{fig:hcm-t}(b), the oscillation appears to have a larger amplitude with a longer and irregular period compared to that of the stretched case. 

For large compressions, $\epsilon=[+0.2\%, +0.6\%]$, well above the critical buckling threshold $\epsilon_c=0.05\%$ of this particular system, the ribbon buckles out of the plane with an amplitude much larger than the unstrained and stretched cases (see Fig.~\ref{fig:schematics_hprofile}(b)). It can be seen from Fig.~\ref{fig:hcm-t}(c) that $\hcm(t)$ of a buckled ribbon behaves like a two level system, and stays buckled either above or below the plane of zero-height with an amplitude much larger than the stretched/unstrained case for a long period of time before it flips to the opposite side (moves to the other minima of a double-well potential). Similar thermally-assisted barrier crossings are  also observed in single-clamped ribbons~\cite{chen2022spontaneous}. This characteristic time, which we will call the escape time (or residence time) $\tau_e$, increases with increasing compression (see Fig.~\ref{fig:hcm-t}(c)). Notice also that when the ribbon stays within a local minimum, it oscillates with a much shorter time scale $\tau_{\rm o}$ than the escape time $\tau_e$, and with a smaller fluctuation amplitude $(\sim0.5a)$ relative to the buckling amplitude $(\sim2a)$.

To summarize, the ribbon oscillates around a single minimum under stretched and unstrained conditions. Beyond the buckling point, however, the ribbon switches between two minima with an escape time $\tau_e$ much larger than the oscillation period inside the wells. By building on these observations and on our mean-field Gibbs free energy Eq.~\ref{eq:GhA}, we will now develop a framework that treats the ribbon midpoint as a Brownian particle confined in a double-well potential of which the strength of the quadratic term is controlled by the external strain (schematically shown in Fig.~\ref{fig:schematics_hprofile}). In the next two sections, we develop a phenomenological theory of the dynamics in the limit of large compression and large stretching energy to explain these observations.
\section{Compressed ribbon dynamics}
\label{sec:compressed-ribbon}
In this section we focus on the dynamics of ribbons under compression above the Euler buckling point. We model the transition from the buckled up state to the down state as a rare event of a transition process over some energy barrier $E_{\rm barrier}$. We begin by discussing the thermally activated transition process of a system in a double-well potential. We then compare the molecular dynamics results with the theoretical predictions~\cite{coffey2012langevin, RevModPhys.62.251}. 
\subsection{Escape time estimated from transition state theory}
The problem of escaping a barrier in a noisy environment, such as a thermal bath, has been studied extensively since the late 1800s, when the well-known Arrhenius form for the escape rate was first formulated based on experimental data~\cite{Arrhenius}
\begin{equation}
    \mathcal{R}=\nu_0 e ^{-E_a/\kBT}, 
\end{equation}
where $\nu_0$ is a prefactor related to an escape frequency and $E_a$ denotes the activation energy. 
Soon after, several theories, summarized in Ref.~\cite{RevModPhys.62.251}, were developed, including Kramers' seminal work~\cite{kramers1940brownian} on incorporating coupling of particles to the heath bath (frictional force), which is missing in the Arrhenius formula. Kramers used a microscopic model of, say, a particle in a nonlinear double-well potential governed by Langevin equations, to formulate the transition rate. The transition rate in the intermediate-to-high damping regime is given by~\cite{kramers1940brownian, coffey2012langevin} 
\begin{equation}
    \mathcal{R}=\left[\left(\frac{\gamma ^2}{4M^2}+\omega_b^2\right)^{1/2}-\frac{\gamma}{2M}\right]\frac{\wo}{2\pi\omega_b}\exp\left[-\frac{E_{\rm barrier}}{\kBT}\right],
    \label{eq:kramers}
\end{equation}
where $E_{\rm barrier}$ is the energy barrier, $\gamma$ is the damping coefficient, $\wo\equiv (U''(x_{\rm min})/M)^{1/2}$ is the angular frequency in the metastable minimum, $\omega_b\equiv (|U''(x_{\rm b})|/M)^{1/2}$ is the angular frequency at the transition (unstable local maximum), $M$ is the particle mass, and $U''(x)$ is the second derivative of a conservative potential $U(x)$. Given that the dynamics of the collective motion, characterized by $\hcm(t)$, of the buckled ribbon and the  effective free energy, with both harmonic and quartic terms, is similar to escape over a barrier, we will first calculate the energy barrier and then discuss the behavior in different temperature regimes. 

Since we work with relatively small strains, we assume a compressible stress $\sigma_{xx}\simeq Y\epsilon$. We define a reduced additional compressive strain relative to critical buckling as $\delta\equiv\frac{\epsilon-\epsilon_c}{\epsilon_c}$. In our previous work we found that the Gibbs free energy can be used to predict thermalized Euler buckling provided that we use the thermally \emph{renormalized} elastic constants $\YT=\Ybare(W_0/\lth)^{-\eta_u}$ and $\kappaT=\kappabare(W_0/\lth)^{\eta}$ whenever $W_0/\lth\gg1$~\cite{hanakata-EML-44-101270-2020}. Following the same approach, we use renormalized elastic constants to calculate $E_{\rm barrier}$, the temperature-dependent critical buckling $\epsilon_c=4\pi^2\kappaT/(\YT L_{\epsilon_c}^2)$, and the maximum height $\hM=\frac{2L_{\epsilon_c}}{\pi}\sqrt{\delta\times\epsilon_c}$. By inserting these renormalized values into Eq.~\ref{eq:kramers}, we obtain the escape time $\tau_e\equiv\mathcal{R}^{-1}$
\begin{align}
    \tau_e=\tau_p\exp
\left[\frac{8\pi^4W_0\kappaT^2\delta^2}{\YT L_{\epsilon_c}^3\kBT}\right].
\label{eq:tau_e}
\end{align}
Here we introduce a prefactor time scale $\tau_p=\left\{\left[\left(\frac{\gamma ^2}{4M^2}+\omega_b^2\right)^{1/2}-\frac{\gamma}{2M}\right]\frac{\wo}{2\pi\omega_b}\right\}^{-1}$, which is the inverse of the prefactor in Eq.~\ref{eq:kramers}. Note that to insure infrequent transitions, $E_{\rm barrier}/\kBT\gg1$ must be satisfied so that we obtain a separation of time scale condition $\tau_e\propto \exp[E_{\rm barrier}/\kBT]\gg\tau_{\rm o}, \tau_{\rm b}$, with $\tau_{\rm o}\equiv 2\pi/\wo$ and $\tau_b\equiv 2\pi/\omega_b$, being the characteristic times at the bottom of the well and at the saddle point, respectively. On using the energy functional with the renormalized elastic parameters (Eqs. \ref{eq:relastics_1}, \ref{eq:relastics_2} and \ref{eq:GhA}) we can directly calculate the renormalized $\tau_{\rm o}$ and $\tau_{\rm b}$, in terms of the areal mass density $\rho$, the ribbon length and the renormalized bending rigidity $\kappaT$,
\begin{equation}
    \tauoT=L_0^2\sqrt{\frac{\rho}{4\pi^2\kappaT\delta}},\quad\taubT=L_0^2\sqrt{\frac{\rho}{2\pi^2\kappaT\delta}}. 
\label{eq:rtau}
\end{equation}
Note that both these times diverge as $\delta \to 0$. Here we use $L_\epsilon\approx L_0$, as we are working with systems with large F\"oppl-von K\'arm\'an number vK number, and $M=\rho W_0L_0$. Our numerical simulations confirm that $L_{\epsilon_c}$ is approximately $L_0$ and weakly
dependent on $T$ as long as $L_0$ is smaller than the persistence length $\lp=\frac{2\kappabare W_0}{\kBT}$ (see Appendix~\ref{sec:lepsilonc}). Note that when $\lp\ll L_0$, the ribbon will behave like a 1D polymer~\cite{kovsmrlj-PRB-93-125431-2016}. In the low-temperature regime $\kappaT\simeq \kappabare$, we recover the $L_0^2$ dependence of the oscillation period $\tau_{\rm o}$, a well-known result for doubly clamped beams~\cite{weaver1991vibration, tilmans1992micro}. 

Upon inserting $\kappaT=\kappabare(W_0/\lth)^\eta$ into Eq.~\ref{eq:rtau} to describe the important intermediate temperature regime, we find $\tauoT, \taubT\propto L_0^z$ with $z=2-\eta/2$. To the best of our knowledge, these deviations in the exponent away from the classical result have not been systematically investigated in experiments $\tau_{\rm o} \propto L_0^2$ scaling in Ref.~\cite{bunch2007electromechanical} and $\tau_{\rm o} \propto L_0$ in Ref.~\cite{chen2009performance}--conclusions which bracket our result $z=2-\eta/2\simeq1.6$, presumably due to relatively larger error bars--nor in numerical simulations. We shall investigate the exponent $z$ and the power law scaling with $T$ numerically in Sec.~\ref{sec:stretched-ribbon}. 
\subsection{Escape time in different temperature regimes}
We first focus on the exponential term, which dominates the behavior for large $\delta$. For convenience in our analysis, we write the term involving $\kappabare^2/\Ybare$ in terms of $\lth^2$.  In the classical low-temperature regime $\lth\gg W_0$ we use the bare elastic constants to obtain 
\begin{align}
\tau_e&=\tau_p\exp\left[\frac{3\pi\delta^2}{8}\left(\frac{W_0}{L_{\epsilon_c}}\right)^3\left(\frac{\lth}{W_0}\right)^2\right].
\label{eq:tau_e_low}
\end{align}
In this regime the ratio between the energy barrier and the thermal energy depends on the cube of the aspect ratio $W_0/L_{\epsilon_c}$ and the square of $\lth/W_0$, yielding the usual Arrhenius-like behavior $\tau_e\propto \exp[E_{\rm barrier}/\kBT]$. 

In the high-temperature regime, however,  we use the renormalized elastic constants $\kappa=\kappabare(W_0/\lth)^\eta$, $Y=\Ybare(W_0/\lth)^{-{\eta_u}}$, as well as the scaling relation $2\eta+\eta_u=2$~\cite{Aronovitz1988}, to obtain the escape time:
\begin{equation}
\tau_e=\tau_p \exp\left[\frac{3\pi\delta^2}{8}\left(\frac{W_0}{L_{\epsilon_c}}\right)^3\right].
\label{eq:tau_e_high}
\end{equation}
Remarkably, and unlike the usual Arrhenius behavior, the exponential term in this case \emph{does not} depend on temperature, but instead depends solely on the geometry, specifically as the cube of the  aspect ratio. 

Now according to Eq.~\ref{eq:kramers}, the prefactor time $\tau_p$ depends on the temperature and strain:
$\tau_p\approx\frac{2\pi\gamma}{M\wo\omega_b}$ for $\gamma/M\gg\omega_b$ and $\tau_p\approx\frac{M}{\gamma}\frac{\kBT}{E_{\rm barrier}}$ for $\gamma/M\ll\omega_b$. (Note that one cannot simply take the limit $\gamma/M\ll\omega_b$ in Eq.~\ref{eq:kramers} to get the \emph{very low} damping regime result. Kramers used a different formulation for this very low damping case~\cite{kramers1940brownian, coffey2012langevin, RevModPhys.62.251}.)

Turning now to the scaling with system size, temperature, and relative compression, we find that in the high-damping regime ($\frac{\gamma}{M}\gg \omega_b$) the prefactor scales as:
\begin{align}
\tau_p &\propto
  \begin{cases}
      L_0^{\,4}\delta^{-1}   & \quad \text{if } W_0/\lth \ll 1\\
    L_0^{\,4-\eta}\delta^{-1}T^{-\eta/2}\sim L_0^{\,3.2}\delta^{-1}T^{-0.4}  & \quad \text{if } W_0/\lth \gg 1.
  \end{cases}
  \label{eq:tau_p_classical}
\end{align}
Here we use $\eta\approx0.8$ and assume some fixed aspect ratio $W_0/L_0$, with $W_0\simeq L_0$, and a fixed ribbon density. With the same assumptions, we can obtain  the prefactor time scale $\tau_p\approx\frac{M}{\gamma}\frac{\kBT}{E_{\rm barrier}}$ at very low damping:
\begin{align}
\tau_p &\propto
  \begin{cases}
      L_0^{\,2}\delta^{-2}T   & \quad \text{if } W_0/\lth \ll 1\\
      \delta^{\,-2} & \quad \text{if } W_0/\lth \gg 1.
  \end{cases}
  \label{eq:tau_p_renormalized}
\end{align}
Thus, apart from the case of low damping and weak (subdominant) renormalization, $\tau_p$ shows either weak or no temperature-dependent behavior. 

Note that we expect Kramers' result to be valid when energy barrier is larger than the thermal energy $\kBT$. By taking the log of the escape time $\tau_e$ (either Eq.~\ref{eq:tau_e_high} or Eq.\ref{eq:tau_e_low}), we see that the $\delta^2$ term (from the energy barrier) dominates the $\log \delta$ term (from $\tau_p$) for large $\delta$. In the next section, we assume that $\tau_p$ is independent of $\delta$ for fitting the extracted escape time $\tau_e$ to either the high temperature result Eq.~\ref{eq:tau_e_high} or the low temperature result Eq.~\ref{eq:tau_e_low}.
Since the exponential term dominates for large $\delta$ and $\tau_p$ is weakly dependent on $T$ for the set of parameters used in our simulations, our analysis suggests that the rather intriguing result that the escape time is controlled only by the geometry when $W_0/\lth\gg1$. In the low-temperature regime, however, we recover the usual Arrhenius behavior with geometry-dependent energy barrier. \section{Molecular dynamics results of compressed ribbons}
\begin{figure}[h]
\centering
\includegraphics[width=8.6cm]{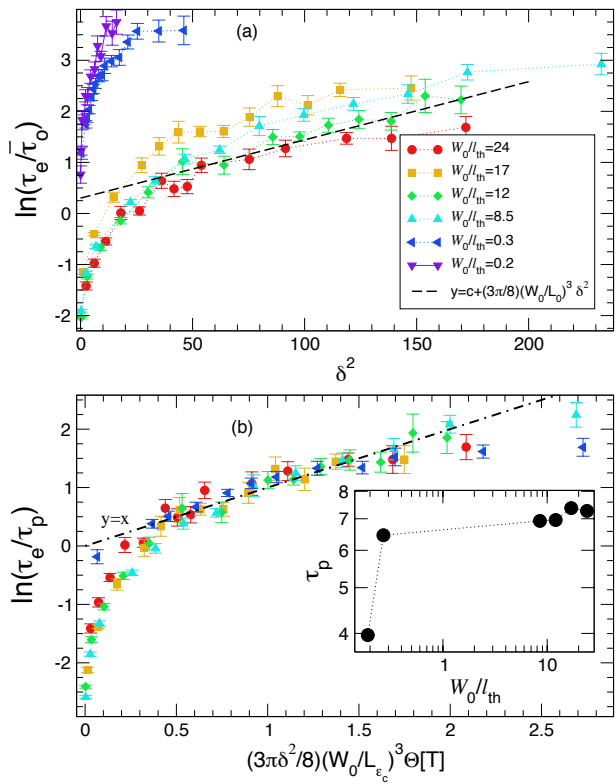}
\caption{(a) Semi-log plot of average escape time, $\tau_e$, in units of the bare oscillation period at zero strain $\overline{\tau_{\rm o}}=L_0^2\sqrt{\frac{\rho}{2\pi^2\kappabare}}$,  as a function of the square of the relative compression, $\delta^2\equiv \left(\frac{\epsilon-\epsilon_c}{\epsilon_c}\right)^2$, at different temperatures such that  $W_0/\lth=[24, 17, 12, 8.5, 0.3, 0.2]$.
Error bars were calculated using the jackknife method~\cite{Young2015}. For large enough compression ($\delta^2\gtrsim50$), so that $E_{\rm barrier}\gg \kBT$, and $W_0/\lth>5$, we see similar slopes approaching $\frac{3\pi}{8}\left(\frac{W_0}{L_0}\right)^3$, shown as a black dashed line. For small values of  $W_0/\lth<0.5$, in contrast, the slopes are higher. (b) Rescaled escape time, $\ln(\tau_e/\tau_p)$, where $\tau_p$ is a fitted prefactor time (inverse attempt frequency), as a function of $\frac{3\pi\delta^2}{8}\left(\frac{W_0}{L_{\epsilon_c}}\right)^3\Theta[T]$, where we use $\Theta[T]=1$ for $W_0/\lth>5$ and $\Theta[T]=(\lth/W_0)^2$ for $W_0/\lth<0.5$. The black dashed-dotted line shows the $y=x$ line, indicating that the data, after appropriate rescaling, agree well with the theoretical prediction. Note that in plot (b) $\tau_p$ is the only adjustable fitting parameter. Inset shows a log-log plot of $\tau_p$ as a function $W_0/\lth$.}
\label{fig:delta2-tau}
\end{figure}
We now turn to molecular dynamics data to test our thermally renormalized stochastic model of a double well potential, beyond the buckling transition. We will use relaxation times extracted from the autocorrelation function $\tau_{\rm AC}$ to approximate the escape time $\tau_e$. Specifically, we calculate the discrete autocorrelation function of the average ribbon height $\hcm$ to quantify ribbon dynamics
\begin{equation}
A_{\hcm}(t_j)=\frac{1}{(n-j)\sigma^2}\sum^{n-j}_{i=1}[\hcm(t_i+t_j)-\mu][\hcm(t_i)-\mu],
\end{equation}
where $n$ is the number of observations in a single simulation run. The time offset is $t_j=j\times \Delta t$ and the sum is over a set of times $t_i=i \times\Delta t$ with $i=1, n-j$. Here and $\mu$ and $\sigma^2$ are the mean and variance of $\hcm$, respectively. In our simulation we choose $\Delta t=10\tau_{\rm MD}$. Given that successive jumps between the up and down states occur at random intervals (see Fig.~\ref{fig:hcm-t}(c)), we expect $A_{\hcm}(t)$ to decay exponentially in time for sufficiently long time $t$ 
\begin{equation}
A_{\hcm}(t)\propto \exp[-t/\tau_{\rm AC}], 
\end{equation}
where $\tau_{\rm AC}$ is the autocorrelation time. 
We average $A_{\hcm}(t)$ over 10 independent runs and fit the data to an exponential function to extract $\tau_{\rm AC}$. While in practice $\tau_{\rm AC}$ may capture more than escape-over-a-barrier dynamics (the longest relaxation time), we expect $\tau_{\rm AC}$ will be dominated by the escape time $\tau_e$, provided we run our simulations long enough to capture at least several rare flipping events; otherwise $\tau_{\rm AC}$ will be on the order of the short-scale relaxation time inside the well. In App.~\ref{sec:three-state-model} we use a different phenomenological theory to extract $\tau_e$ by filtering the up and down states; we still conclude that $\tau_e$ robustly increases with compression and temperature following Eqs.~\ref{eq:tau_e_low} and~\ref{eq:tau_e_high}, at low and high temperature, respectively. 

Fig.~\ref{fig:delta2-tau}(a) shows the rapid increase in $\tau_e$ with increasing compression at a set of temperatures with $W_0/\lth=[24, 17, 12, 8.5, 0.3, 0.2]$. At high temperatures ($W_0/\lth>5$) and sufficiently large $\delta^2$ we see that the slopes, the coefficients in the exponent, are close to $\frac{3\pi^2}{8}\left(\frac{W_0}{L_0}\right)^3$. Remarkably, this high-temperature result indeed indicates the \emph{temperature-independent} ratio of the activation energy to the thermal energy discussed in the previous section.  The slopes in the low-temperature regime ($W_0/\lth\lesssim0.5$), in contrast, increase systematically as the temperature drops. These two very different behaviors are consistent with our earlier analyses based on a thermally renormalized double-well potential for the ribbon height. To further test our theoretical predictions, we fit the high-temperature data ($W_0/\lth>5$) to Eq.~\ref{eq:tau_e_high} and the low-temperature data  ($W_0/\lth<0.5$) to Eq.~\ref{eq:tau_e_low} using \emph{only} $\tau_p$ as an adjustable fitting parameter. As we discussed in the previous section, here we assume that $\tau_p$ is independent of $\delta$ as the exponential of $\delta^2$  dominates for large $\delta$. By rescaling $\tau_e$ with $\tau_p$ and $\delta^2$ with the appropriate temperature and geometrical terms, we are able to collapse all data onto a single curve, as shown in Fig.~\ref{fig:delta2-tau}(b). 

The inset to Fig.~\ref{fig:delta2-tau}(b) shows a log-log plot of the fitted prefactor time $\tau_p$ as a function of $W_0/\lth$. We see that, apart from the lowest temperatures, $\tau_p$ depends weakly on $W_0/\lth$. Fitting only the four high temperature data points ($W_0/\lth>5$) we find $\tau_p={\rm constant}\times T^{0.03}$. This suggests that our data are better described by the low-damping case (see Eqs.~\ref{eq:tau_p_classical} and~\ref{eq:tau_p_renormalized}). Kalmykov et al. provided an exact solution of the correlation time of a Brownian particle in a double-well potential involving special functions~\cite{kalmykov1996exact, coffey2012langevin}. The approximate prefactor of~\cite{kalmykov1996exact, coffey2012langevin}, however, still scales as $1/E_{\rm barrier}\propto\delta^{-2}$, which is still the same as the Kramers' very low damping regime result. Our data in the small $\delta$ regime however do not show such behavior and we do not expect our phenomenological model based on Kramers' result would work for $E_{\rm barrier}\ll \kBT$. Although a more refined theoretical treatment for the prefactor of the escape time of the center-of-mass motion of a double-clamped ribbon is beyond the scope of our current work, we hope to investigate the geometrical and temperature dependencies of this prefactor in the future. 
\section{Stretched and unstrained ribbon}
\label{sec:stretched-ribbon}
We now turn to the dynamics of a ribbon in the regime away from the threshold compressive force needed to produce  the Euler buckling transition, which includes both the stretched and the unstrained cases. Again we approximate the center-of-mass dynamics of the ribbon as a Brownian particle confined in a potential. We first discuss some key results, such as the analytical solution to the positional autocorrelation function within the Brownian particle approximation. The complete derivations of the solution of a  Brownian particle in a harmonic potential can be found in Refs~\cite{zwanzig2001nonequilibrium,coffey2012langevin}, and for completeness we provide key results in the App.~\ref{sec:brownian-harmonic}. 

In the limit of small deflection amplitude, and especially in the large-stretching limit (so that the curvature at the parabolic minimum as a function of $\hcm$ is large),  we can neglect the fourth order term in the potential, retaining only the harmonic term. With this simplification, the equations of motions for the ribbon midpoint become linear,
\begin{align}
  &\frac{d\hcm}{dt}=v\nonumber\\
  &\frac{dv}{dt}=-\frac{\gamma}{M}v-\wo^2\hcm+\frac{1}{M}\xi(t),
\label{eq:1dharmonic}
\end{align}
where $\wo=(k^{\rm eff}/M)^{1/2}$ is the natural frequency, $k^{\rm eff}=\frac{2\pi^2\YT W_0|\epsilon-\epsilon_c|}{L_{\epsilon}}$ is the effective spring constant, $M$ is the  mass, and $\gamma$ is the damping coefficient. The random force $\xi(t)$ is a Gaussian
process with zero mean and $\delta$-function concentrated correlation function. One can solve the Langevin equations in frequency space and obtain the autocorrelation function of $\hcm$ by inverse Fourier transforming the spectral density of the midpoint dynamics $S_{\hcm}\propto|\hcm(\omega)|^2$. The time autocorrelation function of $\hcm$ for the model of Eq.~\ref{eq:1dharmonic} is given by 
\begin{align}
\Ahcm(t)&=\langle \hcm(t')\hcm(t'+t)\rangle\nonumber\\&=\frac{k_BT}{M\wo^2}e^{-\frac{\gamma}{2M}t}\left[\cos \wD t +\frac{\gamma}{2M\wD}\sin \wD t\right],
\label{eq:Ahcm_osc}
\end{align} 
where the natural frequency is renormalized by damping is defined as $\wD=\sqrt{\wo^2-\gamma^2/(4M^2)}$. Thus $A_{\hcm}(t)$ is an
oscillating function with an exponential decay envelope. The damping time $\tau_{\rm damp}\equiv\frac{2M}{\gamma}$ that appears in the exponential prefactor provides a phenomenological description of dissipation in the system. 

\subsection{Molecular dynamics results}
\begin{figure}[h]
\centering
\includegraphics[width=8.6cm]{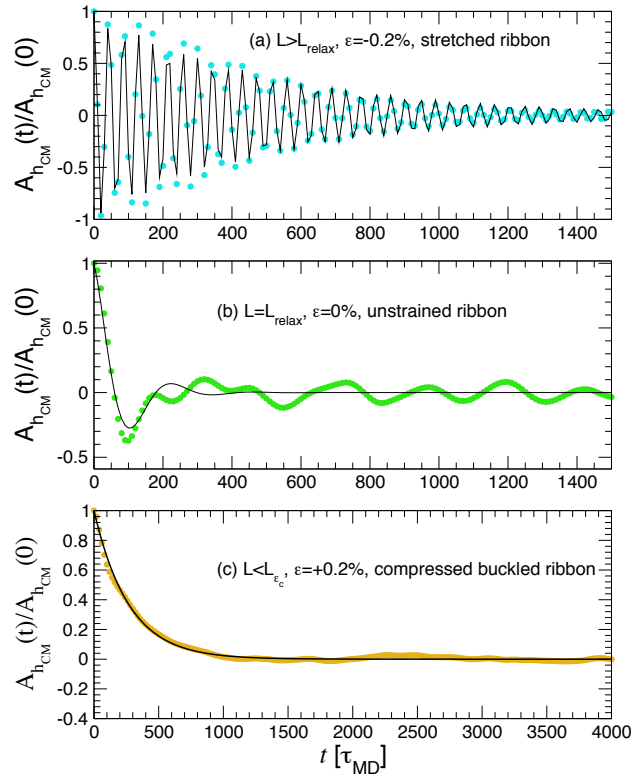}
\caption{Autocorrelation of the midpoint  $\Ahcm$ as a function of time $t$ for (a) stretched, $\epsilon=-2\%$, (b) unstrained, $\epsilon=0\%$, and (c) compressed ribbon, $\epsilon=+2\%$. The last strain is above the buckling threshold, $\epsilon_c=0.05\%$ for our parameter choices. Here the system is thermalized at $W_0/\lth=8.5$, so thermal renormalization of the elastic parameters is important. The circles represent MD data and the black line represents the fitted line. For the stretched and unstrained cases $\Ahcm$ show oscillatory plus exponentially decaying behavior, following Eq.~\ref{eq:Ahcm_osc}. For the buckled case, in contrast, $\Ahcm$ shows only exponential decay.}
\label{fig:A_hcm}
\end{figure}
We now present measurements of time autocorrelation function $A_{\hcm}(t)$ of $\hcm(t)$ from the simulations of a doubly clamped ribbon. Fig.~\ref{fig:A_hcm} shows the average $\Ahcm$ after thermodynamic equilibrium is reached as a function of time $t$ for stretched, unstrained, and compressed ribbons at one representative temperature $\frac{W_0}{\lth}=8.5$ such that thermal fluctuations renormalize the bending rigidity; similar results are found for other parameter choices. For the stretched case $\Ahcm$ oscillates rapidly but decays rather slowly, whereas for the unstrained case $\Ahcm$ oscillates at a lower frequency but decays much faster. When the ribbon is compressed well above the critical buckling threshold, on the other hand, $\Ahcm$ displays a purely exponential  decay. These findings are consistent with the dynamics of $\hcm(t)$ itself shown earlier in Fig.~\ref{fig:hcm-t}.

\begin{figure}[h]
\centering
\includegraphics[width=8.6cm]{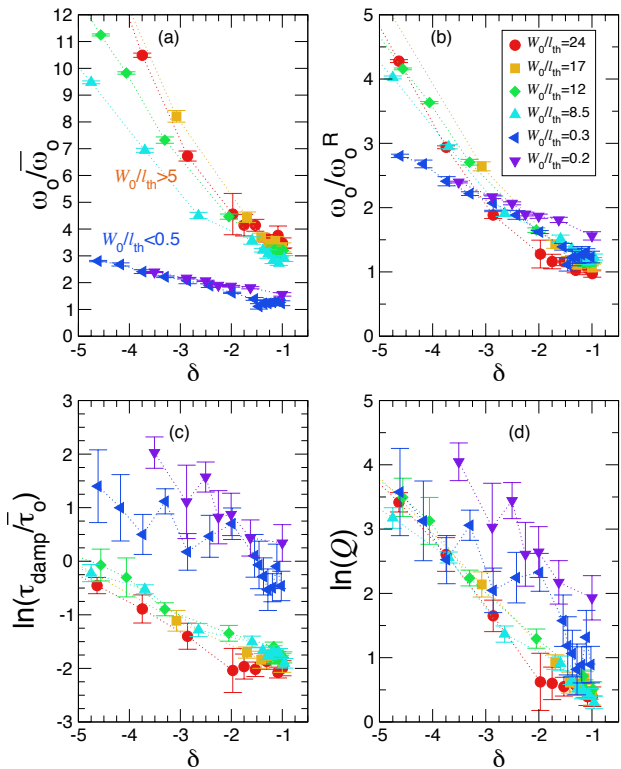}
\caption{(a) The average angular frequency $\wo$ obtained from MD simulations, normalized by its zero-temperature theoretical value at zero strain $\overline{\wo}\equiv\frac{1}{L_0^2}\sqrt{\frac{8\pi^4\kappa}{\rho}}$, as a function of $\delta=\frac{\epsilon-\epsilon_c}{\epsilon_c}$. In the low-temperature regime $W_0/\lth<0.5$ and at zero strain ($\delta=-1$), $\wo/\overline{\wo}\simeq1$. In contrast, in the high- temperature regime $W_0/\lth>5$, $\wo$ at zero strain appears to increase relative to its bare value. (b) Renormalizing  $\wo$ with its thermally renormalized value $\woT$ at zero strain produces a better collapse, suggesting that the stiffening in bending rigidity is reflected in the dynamics.  Log of damping time $\tauD/\overline{\tau_{\rm o}}$ (c) and quality factor $Q=\tauD \wo/2$ (d)  as a function $\delta$. Both $\tauD$ and $Q$ grow with further stretching.}
\label{fig:delta-wo-tauD}
\end{figure}

We next calculate the ribbon resonant frequency $\wo$ from the parameters $\wD$ and $\tauD$ extracted by fitting the data to  Eq.~\ref{eq:Ahcm_osc}. Fig.~\ref{fig:delta-wo-tauD}(a) shows $\wo/\overline{\wo}$ as a function of the relative compression $\delta=\frac{\epsilon-\epsilon_c}{\epsilon_c}$, where  $\overline{\wo}\equiv\frac{1}{L_0^2}\sqrt{\frac{8\pi^4\kappa}{\rho}}$ is the bare natural frequency of the unstrained ribbon without thermal fluctuations ($T=0$, $\delta=-1$). The increase of $\wo$ with increasing tension is consistent with experiments in graphene resonators~\cite{bunch2007electromechanical, chen2009performance}. Notice that, as expected, $\wo/\overline{\wo}\simeq1$ for the two lowest  temperatures ($W_0/\lth<0.5$). In contrast, at high temperatures ($W_0/\lth>5$) $\wo$ increases relative to its zero-temperature value, indicating a stiffening of the bending rigidity. Based on the earlier analysis we can use the predicted renormalized natural frequency at zero strain $\woT\simeq\frac{1}{L_0^2}\sqrt{\frac{8\pi^4\kappaT}{\rho}}$ to rescale the data. Rescaling $\wo$ with its renormalized value $\woT$ yields a better data collapse for not too large $|\delta|$, as shown in Fig.~\ref{fig:delta-wo-tauD}(b). This strategy provides an oscillation measurement route to measuring the stiffening of bending rigidity, a complementary approach to critical buckling measurements~\cite{hanakata-EML-44-101270-2020, morshedifard-JMPS-149-104296-2021}.

Because $\tauD$ also increases with increasing tension, the quality factor of the oscillating ribbon $Q=\tauD \wo/2$ increases with increasing tension. Our simulation data suggest that energy dissipation is reduced for a stretched ribbon, consistent with experimental results for double-clamped graphene~\cite{bunch2007electromechanical, chen2009performance}. 

We also note that we employ the NVT ensemble with Nos\'{e}-Hoover thermostat~\cite{martyna1992nose, martyna1994constant}, which does not have a fixed damping like Langevin dynamics simulations~\cite{schneider1978molecular}. In the NVT ensemble a dynamical term, physically interpreted as a friction, is changing during the approach to thermal equilibration. Once thermal equilibrium at a target temperature is reached, the dynamical friction goes to a finite value and its rate of change vanishes. Given our MD simulations setup, the Brownian particle, embodied in our mean-field description of a thermalized ribbon, is effectively coupled to a thermal bath (thermostat). Consistent with theoretical and numerical investigations of a beam coupled to Nos\'{e}-Hoover thermostat by Louhghalam et al.~\cite{louhghalam2018thermalizing} (see App.~\ref{sec:nose-hoover-beam-theory}), we anticipate an effective damping to occur due to coupling between the doubly clamped ribbon and the thermal bath. This prediction of energy loss, associated with the damping term, is consistent with our MD simulation results. 
\section{Conclusions}
Molecular dynamics simulations of the dynamics of an ultra-thin doubly clamped nanoribbon oscillator reveal rich dynamical behavior. Unlike cantilever geometries, in which stresses relax automatically to produce relatively simple scale-dependent elastic behaviors~\cite{blees2015graphene}, isometrical constraints~\cite{shankar2021thermalized} embodied in double-clamping can lead to an effective tension, a buckling transition and other intriguing phenomena. Thermal fluctuations render the long-wavelength bending rigidity and 2D Young's modulus temperature scale-dependent with important implications for the motion of the center-of-mass height and the two-state nature of the ribbon. The escape time of a ribbon clamped beyond the onset of thermalized Euler buckling grows with increasing compression, as the system must sample two degenerate minima separated by an increasing barrier height. At high temperatures, where thermal fluctuations are significant, the energy barrier for bistable buckled ribbons increases linearly with temperature, thus leading to an approximately temperature-independent Boltzmann factor governing the transition rate. This compensation in the barrier crossing process leads to a transition time in this two-level system that depends only on geometry, in sharp distinction to the low-temperature regime where the escape time increases with the usual Arrhenius-like behavior, $\tau_e\propto e^{E_{\rm barrier}/\kBT}$. 

For a stretched ribbon we find that the natural angular frequency $\wo$ and the quality factor $Q$ increases with increasing tension, consistent with experiments~\cite{bunch2007electromechanical, chen2009performance, zande2010large}. Our theoretical work indicates that in the high-temperature regime the oscillation period scales with ribbon size $L_0$ and temperature $T$ as $\tauoT \propto \frac{1}{\woT}\sim L_0^{(2-\eta/2)}T^{-\eta/4}$. This scaling with ribbon size $L_0$ suggests that thermalized nanoribbons close to the buckling transition, so that the ribbons are relaxed, behave as a system with a dynamical critical exponent of $z=2-\eta/2=1.6$, assuming that the static critical exponent $\eta$ is 0.8. Several experiments  on doubly clamped graphene ribbons have shown either $L_0^{2}$~\cite{bunch2007electromechanical} or $L_0$~\cite{chen2009performance} scaling of the inverse of the natural frequency. These experimental results, of limited precision, bracket the exponent  $z=1.6$ found here. This scaling behavior could be tested by computational and experimental work that systematically varies the system size whilst ensuring vanishing tension. 

In this context, we mention a very recent theoretical investigation of the dynamics of \emph{free-standing} graphene~\cite{granato2022dynamic}, which is also motivated by experimental investigations~\cite{ackerman2016anomalous}. Granato et al. argue that the time behavior of the mean-square displacement of height fluctuations, $\langle \Delta h(t)^2 \rangle$, at long and intermediate times, should not depend on the microscopic length. This argument, together with the scaling of elastic membranes, leads to $\langle \Delta h(t)^2 \rangle \sim t^{\frac{\zeta}{1+\zeta}}$ with $\zeta=(1-\eta/2)$ being the roughening critical exponent, a static equilibrium quantity. Further dimensional analysis by Granato et al. suggests that the subdiffusive time scale of the mean-square displacement has the form $\tau\sim L_0^{2(1+\zeta)}\sim L_0^{4-\eta}$. Our work concerns the dynamical exponent of different physical quantities: (i) the characteristic oscillation time of the midpoint inside a minima $\tau_{\rm o}\sim L_0^{2-\eta/2}$ and (ii) the characteristic prefactor time scale of the escape time with $\tau_p\sim L_0^{\,4-\eta}$ in the high-damping regime and $\tau_p$ is independent of system size in the low-damping regime. 

Our simulation results confirm that $\wo$ increases with increasing temperature due to stiffening in bending rigidity, which is consistent with our theoretical model. Several experiments have shown that the natural angular frequency $\wo$ and the quality factor $Q$ of graphene resonators indeed increases with decreasing temperature~\cite{chen2009performance, zande2010large}.  In contrast, other experiments on graphene resonators showed that the natural frequency increases with increasing temperature~\cite{oshidari2012high, ye2018electrothermally}. Both sets of experiments conclude that frozen strains due to cooling/heating cycles could play an important role in altering the resonant frequency. The strain, however, is not directly controlled in those experiments. Our simulations, in contrast, allow us to examine the temperature dependence of $\wo$ while keeping the relative compression constant across different temperatures. We show that, for a fixed reduced stretching strain, $\wo$ increases with temperature according to $\wo\sim T^{\eta/4}$, due to bending stiffening above the temperature at which thermal renormalization effects take place. Another challenge requiring further study is the temperature-dependence of $Q$ and $\wo$ where the energy loss due to boundary effects, such as  imperfect clamping and different thermal expansions across different materials which present in physical experiments~\cite{mohanty2002intrinsic, imboden2014dissipation}, is taken into account. Future investigations might include simulating a ribbon adhered to a substrate via an attractive microscopic potential as opposed to the perfect clamping condition imposed in our current work. 

In summary, we have investigated the dynamics of the midpoint  of doubly clamped nanoribbons at a wide range of temperatures. This work suggests that dynamical measurements may be used as an alternative way to study the unusual thermal renormalization of the underlying elastic constants.  We hope that our work will encourage  theoretical and experimental investigations of  the non-trivial dynamical exponents of atomically thin ribbons of, e.g. graphene and MoS$_2$. From a practical standpoint, our findings are important for 
predicting the response of nanoactuators and non-linear mechanical nanoresonators operating at wide range of temperature and strain conditions. 

\begin{acknowledgments}
P.Z.H. and D.R.N. acknowledge support through NSF Grant No. DMR-1608501 and via the Harvard Materials Science Research and Engineering Center, through NSF Grant No. DMR-2011754. We also thank the KITP program, ``The Physics of Elastic Films: From Biological Membranes to Extreme Mechanics," supported in part by the National Science Foundation under Grant No. NSF PHY-1748958. D.Y. acknowledges support from Ministerio de Economía y Competitividad (MINECO) and Agencia Estatal de Investigación (Spain) through grant no. PGC2018-094684-B-C21, partially funded by the European Regional Development Fund (FEDER, European Union). HOOMD simulation input scripts and other codes are available at \url{https://github.com/phanakata/statistical-mechanics-of-thin-materials/}. We thank Roberto Valenzuela, Suraj Shankar, Daniel Lopez, Richard Huang and Abigail Plummer for helpful discussions. P.Z.H also thanks Harold Park and Jin-Wu Jiang for helpful discussions on energy dissipation in nanomechanical systems. 
\end{acknowledgments}

%\end{document}
\newpage

\appendix
In these Appendices we provide detailed derivations, supplementary molecular dynamics data, and a complementary phenomenological theory describing the dynamics that are not included in the main text. 
%\tableofcontents
\section{Brownian particle in 1D harmonic potential}
\label{sec:brownian-harmonic}
\begin{figure}
\includegraphics[width=8.6cm]{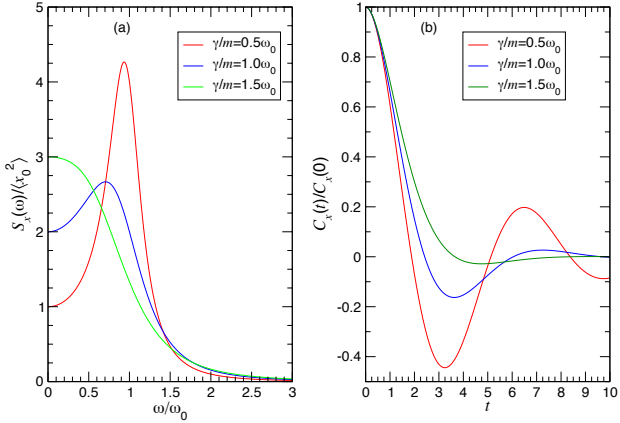}
\caption{(a) Normalized spectral density
  $S_{\hcm}(\omega)/\langle x_0^2\rangle$ as a function of frequency
  $\omega$ and (b) correlation of position $C_{x}(t)$ as a
  function of time $t$ for different values of $\gamma/m$.}
\label{fig:S_x_and_C_x}
\end{figure}
We consider a Brownian particle of mass $m$ allowed to move in the $x$ direction and confined in a harmonic
potential $V(x)=kx^2/2$. This model is used to approximate the center of mass $\hcm$ of an unbuckled ribbon well below the threshold for the Euler buckling transition, although the fourth order quartic term will lead to some corrections to the results in this Appendix (see our our coarse-grained Gibbs free
energy). The equations  of motions are given by
\begin{align}
  &\frac{dx}{dt}=v \label{eq:langevin_1}\\
  &\frac{dv}{dt}=-\frac{\gamma}{m}v-\omega_0^2x+\frac{1}{m}\xi(t),
  \label{eq:langevin_2}
\end{align}
where $\omega_0^2=k/m$ defines the oscillator frequency associated with the harmonic potential at $x=\hcm=0$ for the ribbon.  The random force $\xi(t)$ is a Gaussian
process with zero mean and correlation function proportional to
$\delta$-function
\begin{equation}
\langle\xi(t)\rangle=0,\quad \langle \xi(t)\xi(t')\rangle=2\gamma k_BT\delta(t-t').
\end{equation}
Upon Fourier transforming the Langevin  equations(Eqs.~\ref{eq:langevin_1} and \ref{eq:langevin_2}) to the frequency domain
\begin{align}
  &-i\omega x(\omega)=v(\omega)\\
  &-i\omega v(\omega)=-\frac{\gamma}{m}v(\omega)-\omega_0^2x(\omega)+\frac{1}{m}\xi(\omega),
\end{align}
and upon solving the equations above we obtain
\begin{equation}
x(\omega)=\frac{1}{m}\frac{\xi(\omega)}{\omega_0^2-\omega^2-i\frac{\gamma}{m}\omega}.
\end{equation}
It is useful to study the amplitude of $x(t)$ in frequency space to
understand the dynamics.  A closely related and commonly measured quantity
in signal processing and studies of Brownian motion
is the spectral density $S_x(\omega)\propto |x(\omega)|^2$:
\begin{align}
S_x(\omega)&=\frac{1}{m}\frac{\langle|\xi(\omega)|^2\rangle}{|\omega_0^2-\omega^2-\frac{\gamma}{m}i\omega|^2}\\\nonumber
&=\frac{2\gamma k_BT}{m^2\left[(\omega_0^2-\omega^2)^2+\frac{\gamma^2}{m^2}\omega^2\right]}.
\label{eq:S_x}
\end{align}
From the equipartition theorem we expect
$\omega_0^2\langle x_0^2\rangle/2=k_BT/2$. We can normalize
$S_x(\omega)$ by inserting $\langle x_0^2\rangle=k_BT/(m\omega_0^2)$. We plot
$\frac{S_x}{k_BT/(m \omega_0^2)}$ (Eq.~\ref{eq:S_x}) as a function of $\omega$ for a
fixed $\omega_0=1$ and different values of $\gamma/m, k_BT$ in
Fig.~\ref{fig:S_x_and_C_x}(a). Similarly, by inverting the Fourier transform, we can calculate the
position autocorrelation function 
\begin{align}
C_x(t)&=\frac{1}{2\pi}\int^{\infty}_{-\infty}d\omega e^{-i\omega t}S_x(\omega)\\
&=\frac{\gamma k_BT}{\pi m^2}\int^{\infty}_{-\infty}d\omega e^{-i\omega t}\frac{1}{\left[(\omega_0^2-\omega^2)+\frac{\gamma^2}{m^2}\omega^2\right]}\\
&=\frac{k_BT}{m\omega_0^2}e^{-\frac{\gamma}{2m}t}\left[\cos \omega_1t +\frac{\gamma}{2m\omega_1}\sin \omega_1t\right],
\end{align} 
where $\omega_1=\sqrt{\omega_0^2-\gamma^2/(4m^2)}$ is the damped natural frequency. $C_x(t)$ is an
oscillating function with exponential decay. The solution for the $x$-autocorrelation
functions $C_x(t)$ for different values of $\gamma/m$ are plotted in
fig.~\ref{fig:S_x_and_C_x}(b).
\section{Temperature behavior of critical buckling length}
\label{sec:lepsilonc}
\begin{figure}
\includegraphics[width=8.6cm]{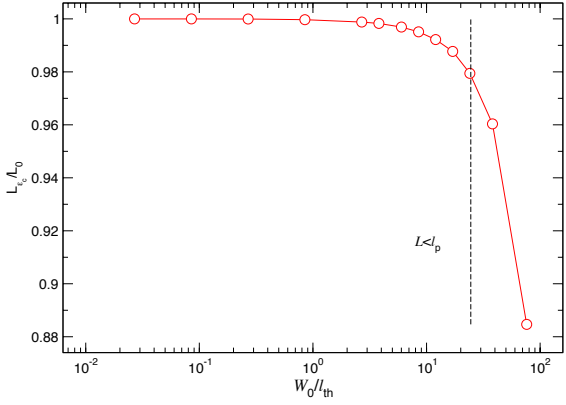}
\caption{The projected length $L_{\epsilon_c}$ at the critical buckling strain as a function of $W_0/\lth$. $L_{\epsilon_c}$ is weakly dependent on temperature when $L<\lp$. }
  \label{fig:lth-Lc}
\end{figure}
For a large vK number, the critical buckling strain $\epsilon_c\propto \kappabare/\Ybare L_0^2$ is generally very small.  Hence, the projected critical buckling length should be close to the undeformed zero-temperature (rest) length $L_0$. From MD simulations we indeed find that $L_{\epsilon_c}$ weakly depends on $T$ as long as the ribbon length is smaller than the persistence length $\lp=\frac{2\kappabare W_0}{\kBT}$, as shown in Fig.~\ref{fig:lth-Lc}. 
\section{Three-state model and residence time estimation}
\label{sec:three-state-model}
\begin{figure}
\includegraphics[width=8.6cm]{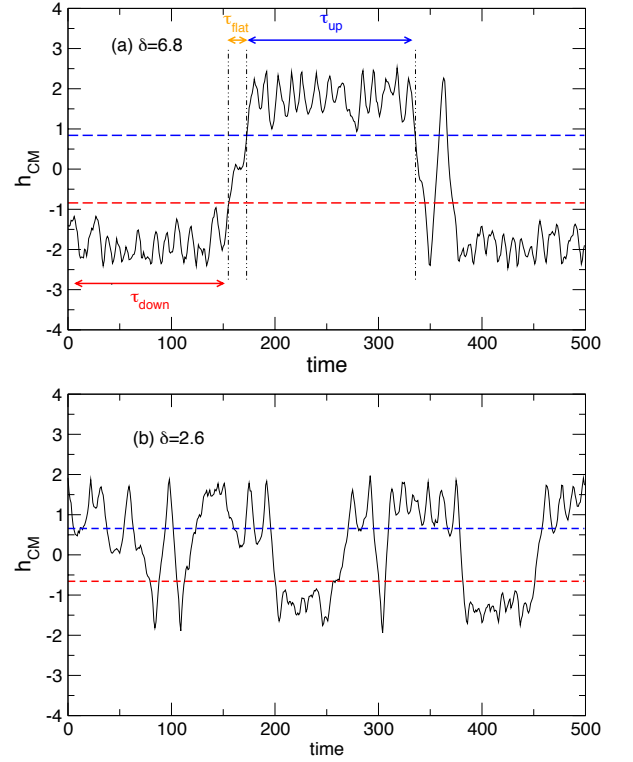}
\caption{Midpoint $\hcm$ as a function of time in units of $10\tau_{\rm MD}$ at a strain (a) above the buckling transition with $\delta=6.8$ and (b) above but closer to the buckling transition with $\delta=2.6$. Well above the buckling transition, the ribbon spends most of its time in either the up or down state. In contrast, close to the buckling point the ribbon transitions from the up to the down state more frequently, and so spends its time in the up, down, and flat state more equally. The system shown here is thermalized at $W_0/\lth\sim8.5$.}
\label{fig:hcm-t-filter}
\end{figure}

\begin{figure}
\includegraphics[width=8.6cm]{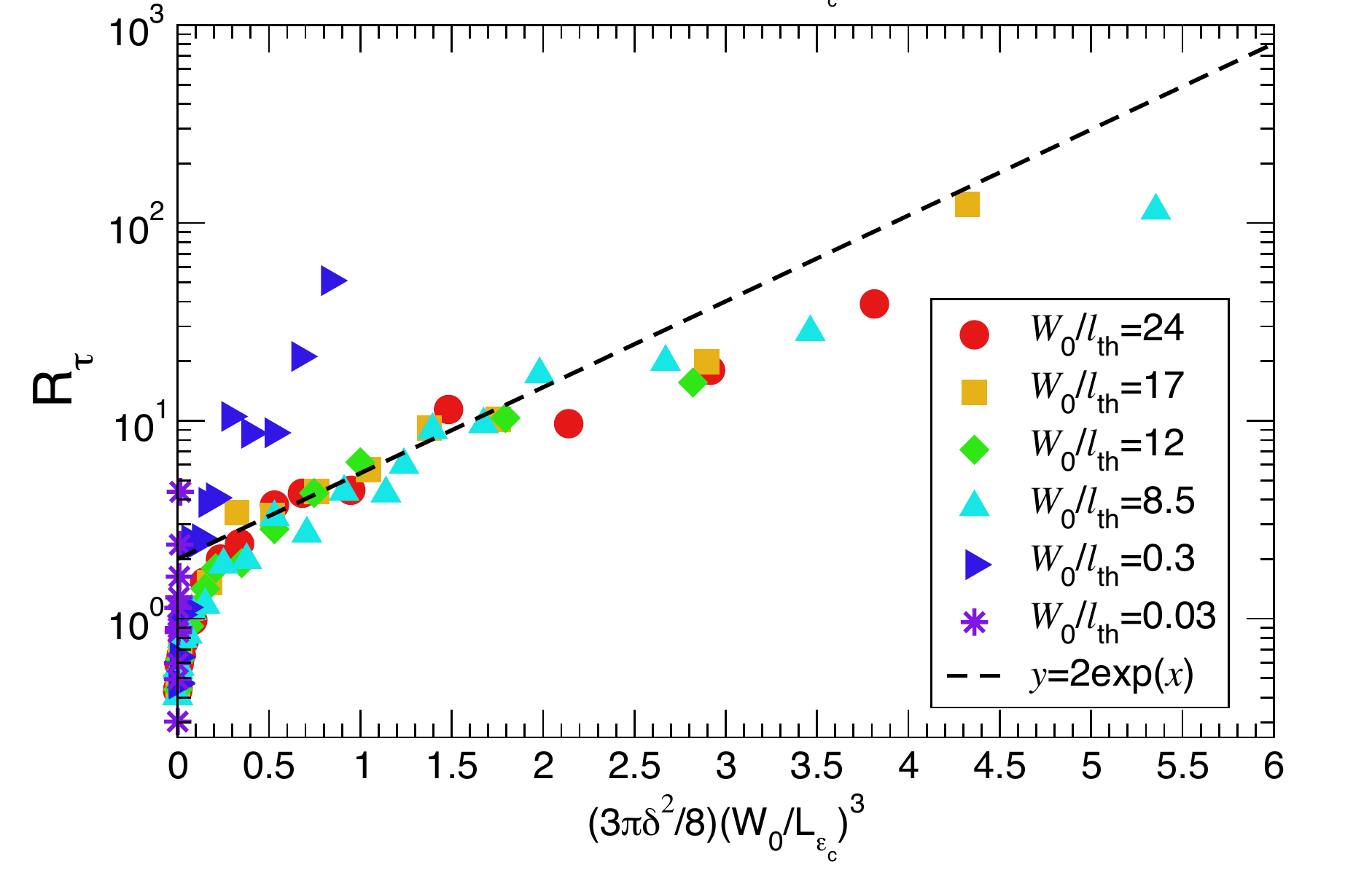}
\caption{The time ratio $R_\tau$ as a function
  $3\pi\delta^2/8(W_0/\lth)^3$. The slope is close to one, consistent with the theoretical prediction.  }
  \label{fig:Rtau}
\end{figure}
In the main text we use Kramers' result to describe the escape time. Here, we develop a three-state model as a complementary theory to describe the ribbon dynamics above the critical buckling. Suppose that we only have three possible states (Up, Down, and Flat) with energies $E[\pm |h_{\rm CM}|]=-E_{\rm barrier}$ and $E[0]=0$. The probability of being in a given state is proportional to the Boltzmann factor, and the probability of being in the up state is given by
\begin{equation}
  P(+h_{\rm CM}) = \frac{\exp [E_{\rm barrier}/\kBT]}{1+2 \exp[E_{\rm barrier}/\kBT]}.
\end{equation} 
In simulations we can relate this probability to time as $\sum P(E)=1$ and $\sum \tau(E)/T=1$ in the limit $T\rightarrow\infty$. We can then estimate the ratio between the total time in the up- and down-states and the time in the flat state to be
\begin{equation} 
  R_\tau=\frac{\sum\tau_{\rm up}+\tau_{\rm down}}{\sum\tau_{\rm flat}}\propto 2 \exp[ E_{\rm barrier}/k_{\rm B}T],
\end{equation}
where $E_{\rm barrier}$ is given in the main text. 

In two different temperature regimes separated by thermal length $\lth$,  the time ratio $R_\tau$ is given by
\begin{equation}
R_{\tau}\propto
  \begin{cases}
         & \exp\left[\frac{3\pi\delta^2}{8}\frac{W_0\lth^2}{L_{\epsilon_c}^3}\right]\text{if } W_0\ll \lth,\\
     & \exp\left[\frac{3\pi\delta^2}{8}\frac{W_0^3}{L_{\epsilon_c}^3}\right]\text{if } W_0\gg \lth,
  \end{cases}\label{eq:RtauR}\\
\end{equation}
We first test this relation for systems with $W_0>\lth$ (semi-flexible regime). We expect $\log(\frac{\tau_{\rm up}+\tau_{\rm down}}{\tau_{\rm flat}})= {\rm slope}\times\delta^2 + c$, where the slope is obtained from theory $\left(\frac{W_0}{L_{\epsilon_c}}\right)^3\frac{3\pi}{8}\sim0.01$.  To extract $\tau$ we use a height threshold $h_c=h_{\rm max}/3$ and define an up- or down-state whenever $|h_{\rm CM}|>h_c$. Fig.~\ref{fig:hcm-t-filter} shows the midpoint $h_{\rm CM}$ as a function time for a ribbon well above the buckling transition and close to the buckling transition. Well above the buckling transition, the ribbon spends most of its time in either the up or down state. Close to the buckling transition, in contrast, the ribbon switches from the up to the down state more frequently, and so the ribbon spends its time in the up, down, and flat state more
equally. Fig.~\ref{fig:Rtau} shows the time ratio
$R_\tau=\frac{\sum\tau_{\rm up}+\tau_{\rm down}}{\sum\tau_{\rm flat}}$ as a function of $3\pi\delta^2/8 (W_0/\lth)^3$. Close to the buckling transition $\frac{\tau_{\rm up}+\tau_{\rm down}}{\tau_{\rm flat}}\sim 2$, which suggests that all three states are equally probable. Since  $E_{\rm barrier}=0$ at the transition, all three states are equally probable. From simulations we find that the slope is close to the analytical prediction.  Note we could model the buckling problem as two states only (Up, Down). One can compute the cumulative probability distribution of the residence times and calculate the integrated survival time as a measure of the escape time. As shown in Ref.~\cite{bhabesh2018statistical}, the integrated survival time $\tau_{\rm surv}$ is proportional to the autocorrelation time ($\tau_{\rm AC}\sim0.5\tau_{\rm surv}$) as the autocorrelation time is related to the slowest mode of interest. This three-state model is used as a complementary theory showing how the activation energy becomes renormalized for $W_0/\lth\gg1$, with the advantage that no prefactor is needed.   

\section{Nos\'{e}-Hoover Beam Theory}
\label{sec:nose-hoover-beam-theory}
In the main text we developed a mean-field model that treats the many connecting nodes of a ribbon as a one-dimensional problem. This problem is equivalent to beam theory, however, with renormalized elastic constants. Our molecular dynamics simulations were carried in a canonical (NVT) ensemble where number of particles $N$, volume $V$ and temperature $T$ are fixed. Within this ensemble we used the Nos\'{e}-Hoover thermostat~\cite{martyna1994constant, martyna1992nose} implemented in HOOMD-blue~\cite{anderson2020hoomd}. Thus, we need to add a thermal bath to our mean-field model in order to explain the observed quantities, such as height oscillations. In this appendix we provide derivations of the equation of motion for a beam coupled to a thermal bath, first derived in Ref~\cite{louhghalam2018thermalizing}. Note that here we followed the notation in Ref~\cite{louhghalam2018thermalizing}.

In a microcanonical (NVE) ensemble number of particles $N$, volume $V$ and energy are conserved. The Lagrangian, the difference between the kinetic and the potential energy, of a beam in the absence of an external force is given by
\begin{equation}
\mathcal{L_{\rm beam}}=\int\left[\frac{1}{2}\rho A\dot{h}^2-\frac{1}{2}EI(h'')^2\right]dx,
\label{eq:Lbeam}
\end{equation}
where $\rho$ is the density, $h(x)$ is the height deflection, $A$ is the beam cross section, $EI$ is the bending stiffness, $h'=\partial h/\partial x$ and $\dot{h}=\partial h/\partial t$. Note that the quartic term is not included, unlike our mean field model for a clamped ribbon. Using the Euler-Lagrange equation resulting from 
Eq.~\ref{eq:Lbeam}, we obtain the equation for undamped motion of a
beam in the NVE-ensemble
\begin{align}
&-\frac{\partial}{\partial t}\left(\frac{\partial \mathcal{L}_{\rm beam}}{\partial \dot{h}}\right)+\frac{\partial ^2}{\partial x^2}\left(\frac{\mathcal{L}_{\rm beam}}{\partial h''}\right)=0\\
& \Rightarrow\rho A\ddot{h}+\frac{\partial ^2}{\partial x^2}EIh''=0. \label{eq:undamped}
\end{align}
In a canonical ensemble the system, which in this case is the beam, is in contact with a thermal bath with a reference temperature $T_{\rm ref}$. The extended Lagrangian is $\Lex=\Lbeam+\Lbath$. In the Nos\'{e}-Hoover thermostat, a fictitious mass $Q>0$ of dimension $ML^2$ and its velocity $\zeta$ of dimension time$^{-1}$ are introduced~\cite{martyna1992nose, martyna1994constant}. The bath potential energy is $RT_{\rm ref}\ln(s)$, with $s$ being the generalized coordinate and $R$ the product of the Boltzmann constant and the number of degrees-of-freedom. The generalized coordinate $s$ and the velocity $\zeta$ are related by
\begin{equation}
\zeta=\frac{ds}{d\tau}\quad s=\frac{d\tau}{dt}, 
\end{equation}
where $\zeta$ determines the heat exchange between the beam and the bath and $s$ is the stretch in time between the time of the beam, $t$, and the time of the bath $\tau$.  The bath Lagrangian is
\begin{equation}
\Lbath=\frac{Q}{2}\zeta^2-RT_{\rm ref}\ln(s).
\end{equation} 
Before moving further we first relate the time derivatives
\begin{align}
&\frac{\partial s}{\partial t}=\zeta s \label{eq:change1}\\
&\dot{h}=s\frac{\partial h}{\partial \tau}\label{eq:change2}\\
&\ddot{h}=s\frac{\partial}{\partial \tau}\left(s\frac{\partial h}{\partial \tau}\right)=s^2\frac{\partial ^2h}{\partial \tau^2}+s\zeta\frac{\partial h}{\partial \tau}\label{eq:change3}
\end{align}
By change of variables we can write the extended Lagrangian in the extended time scale $s$
\begin{equation}
\Lex=\int^{L}_0\left[\frac{\rho A s^2}{2}\left(\frac{\partial h}{\partial \tau}\right)^2-\frac{EI}{2}\left(\frac{\partial^2 h}{\partial x^2}\right)^2\right]dx+\frac{Q\zeta^2}{2}-RT_{\rm ref}\ln(s).
\end{equation}
As earlier, we obtain the equation of motion by using Euler-Lagrange equation
\begin{align}
&-\frac{\partial}{\partial \tau}\left(\frac{\partial \Lex}{\partial (\partial h/d\tau)}\right)+\frac{\partial ^2}{\partial x^2}\left(\frac{\Lex}{\partial h''}\right)=0\\
& \Rightarrow\rho A\left(s^2\frac{\partial^2 h}{\partial \tau^2}+2\zeta s \frac{\partial h}{\partial \tau}\right)+\frac{\partial ^2}{\partial x^2}EIh''=0.
\end{align}
 We can use Eqs.~\ref{eq:change1}, ~\ref{eq:change2} and ~\ref{eq:change3} to rewrite the equation of motion in the real time ($t$)
\begin{equation}
\rho A\ddot{h}+\rho A \zeta\dot{h}+\frac{\partial ^2}{\partial x^2}EIh''=0.
\end{equation}   
Notice that this is similar to the undamped case Eq.~\ref{eq:undamped} but now we have a new damping term $\propto \zeta \dot{h}$, which is similar to the friction term in Langevin dynamics. Using the Euler-Lagrange equation we  obtain the equation for the evolution of $\zeta$:
\begin{equation}
    \dot{\zeta}=\frac{d}{dt}\left(\frac{\partial \ln(s)}{\partial t}\right)=\frac{RT_{\rm ref}}{Q}\left(\frac{T(t)}{T_{\rm ref}}-1\right).
\end{equation}
As $T/T_{\rm ref}\rightarrow1$ the friction term $\zeta$ tends to a constant, indicating equilibrium. 

In summary, we have shown that coupling a beam to a thermal
bath results in an effective damping. The consequences of the
mean-field theory with coupling to the bath is consistent with our simulation, given that we use Nos\'{e}-Hoover thermostat for the NVT molecular dynamics simulations. This damping (energy loss) is observed in our simulation data, which is characterized by a decaying oscillation of the positional correlation function in the stretched ribbon case and in a purely decaying behavior of the positional correlation function for the buckled case.

\bibliography{dynamics_buckling}
\end{document}